# The valence and Rydberg states of difluoromethane: a combined experimental vacuum ultraviolet spectrum absorption and theoretical study by *ab initio* configuration interaction and density functional computations.

Michael H. Palmer,[1,a] Søren Vrønning Hoffmann,[2,b] Nykola C. Jones,[2,b] Marcello Coreno,[3,b] Monica de Simone,[4,b] Cesare Grazioli.[3,b]

[1] *School of Chemistry, University of Edinburgh, Joseph Black Building, David Brewster Road, Edinburgh EH9 3FJ, Scotland, UK*

[2]*Department of Physics and Astronomy, ISA, Aarhus University, Ny Munkegade 120, DK-8000 Aarhus C, Denmark*

[3] *ISM-CNR, Montelibretti, c/o Laboratorio Elettra, Trieste, Italy*

[4] *CNR-IOM Laboratorio TASC, Trieste, Italy*

[a)] Email: *m.h.palmer@ed.ac.uk*: Telephone: +44 (0) 131 650 4765

[b)] *Electronic addresses: vronning@phys.au.dk; nykj@phys.au.dk; marcello.coreno@elettra.eu; desimone@iom.cnr.it; cesare.grazioli@elettra.eu*

## ABSTRACT

The vacuum ultraviolet spectrum (VUV) for $CH_2F_2$ from a new synchrotron study, has been combined with earlier data, and subjected to detailed scrutiny. The onset of absorption, band I, and also band IV, are resolved into broad vibrational peaks, which contrast with the continuous absorption previously claimed. A new theoretical analysis, using a combination of time dependent density functional theory (TDDFT) calculations and complete active space self-consistent field (CASSCF), leads to a major new interpretation. Adiabatic and vertical excitation energies (AEE and VEE), evaluated by these methods, are used to interpret the spectra in unprecedented detail using theoretical vibronic analysis. This includes both Franck-Condon (FC) and Herzberg-Teller (HT) effects on cold and hot bands. These results lead to re-assignment of several known excited states, and identification of new ones. The lowest calculated AEE sequence for singlet states is: $1^1B_1 \sim 1^1A_2 < 2^1B_1 < 1^1A_1 < 2^1A_1 < 1^1B_2 < 3^1A_1 < 3^1B_1$. These, together with calculated higher energy states, give a satisfactory account of the principal maxima observed in the VUV spectrum. Basis sets up to quadruple zeta valence with extensive polarization (QZVPPD) are used. The diffuse functions within this type of basis generate both valence and low-lying Rydberg excited states. The optimum position for the site of further diffuse functions in the calculations of Rydberg states is shown to lie on the H-atoms. The routine choice on the F-atoms, is shown to be inadequate for both *CHF₃* and *CH₂F₂*. The lowest excitation energy region has mixed valence and Rydberg character. TDDFT calculations show that the unusual structure of the onset arises from the near degeneracy of $1^1B_1$ and $1^1A_2$ valence states, which mix in symmetric and antisymmetric combinations. The absence of fluorescence in the 10.8 to 11 eV region contrasts with strong absorption. This is interpreted by the $2^1B_1$ and $1^1A_1$ states where no fluorescence is calculated for these two states, which are only active in absorption. The nature of the two states, $1^1B_1$ and $2^1B_1$ are fundamentally different, but both are complex owing to the presence of FC and HT effects occurring in different ways. The two most intense bands, close to 12.5 and 15.5 eV, both contain valence states as expected; the onset of the 15.5 eV band shows a set of vibrational peaks but the vibration frequency does not correspond to any of the PES structure, and is clearly valence in nature. The routine use of photoelectron spectral (PES) footprints to detect Rydberg states in VUV spectra, is shown to be inadequate. The combined effects of FC and HT in the VUV spectral bands lead to additional vibrations when compared with the PES.

## I. INTRODUCTION.

Difluoromethane ($CH_2F_2$) is one of the hydrofluorocarbons (*HFC-32*); it has been used as an alternative coolant to the ozone destroying chlorofluorocarbons (*CFCs*, also known as *Freon's*) in air-conditioning and refrigerating systems. However, $CH_2F_2$ and related chemicals also are being phased out, since its global warming potential is approximately 675 times greater than that of $CO_2$ itself.[1,2] *HFC-32* is an ozone-depleting substance indirectly; photolysis of ozone produces O($^1$D), which is an important atmospheric oxidant of $CH_2F_2$.[3] These issues make it important to accurately interpret as much spectral information for $CH_2F_2$ as possible. Currently, availability and transportation of $CH_2F_2$ is restricted in Europe.

We have recently reported[4] a rigorous re-analysis of the photoelectron spectrum (PES) for $CH_2F_2$. The present study makes use of these results in a detailed re-analysis of the vacuum ultraviolet (VUV) spectrum, alongside rigorous theoretical methods of study.

As expected for small molecules, there are several previous VUV spectral studies of both $CH_2F_2$ and $CH_2Cl_2$ which are relevant here.[5-12] Early reports into the nature of the VUV spectrum for $CH_2F_2$, include four discrete absorption regions below 13 eV, being attributed to Rydberg rather than valence states.[5-8] No information on electronic states was extracted from 100 eV electron impact (EI)[11] studies, where the VUV and EI profiles are effectively identical. Comparison[9] of the VUV absorption with fluorescence spectra for $CH_2F_2$ shows a strong similarity for the first two bands, which are centered on 9.28 and 10.29 eV. In contrast, the complex VUV band between 11 and 12 eV is almost completely absent in fluorescence. Above this energy, there is considerable difference between the absorption and fluorescence spectra as expected; we discuss this below. Shastri *et al.*[10,11] assigned numerous Rydberg states for $CH_2F_2$ in the region 8.3 to 11.8 eV, using ionization energies determined by Pradeep and Shirley[13] and Brundle *et al.*[14] These VUV assignments have been questioned,[9] and are also addressed below.

Each of the molecules $CH_2F_2$, $CH_2Cl_2$ and $CD_2Cl_2$ (D = $^2H$) are isoelectronic in the valence shell. Continuous absorption for the chloro-compounds at the spectral onset,[10,11] has been attributed[14] to overlapping electronic states, rather than the single state previously assumed.[5-9] The 7 to 12 eV VUV absorption of the series $CH_kCl_{4-k}$ ($k$ = 0 to 3), has been generally attributed to excitation of non-bonding electrons on chlorine, and this was consistent with several studies for $CH_2F_2$.[5-11] However, in a theoretical study of the series $CH_kF_{4-k}$ ($k$ = 0 to 3), Edwards and Raymonda[7] concluded that the first absorption in $CHF_3$, attributed to a Rydberg state, should be associated with the *C-H* unit alone. The first absorption band relating to the $CF_3$ group was thought to lie significantly higher, while the lowest IE for $CHF_3$ is close to that for atomic hydrogen.[7] This implied that the Rydberg excitation for $CHF_3$ was essentially localized in the *C-H* bond,[7] rather than in orbitals on the *C* or *F* atoms. We will offer support for this view below.

In order to discuss the electronic state assignments for a wider energy range, we combine our new VUV spectrum with data for the 9 to 21 eV range, kindly provided by Seccombe *et al*.[9] Thus, we use the same VUV data for the region above 10.8 eV, but reanalyze it in greater detail. We adopt their nomenclature[9] where the series of broad bands in the VUV spectrum between 8 and 18 eV are denoted as bands I through VIII. This process was greatly assisted by 'subtraction' of strong broad structures, using a mathematical 'fit' to local areas of the spectrum, and processing the resulting regular residuals separately. Our VUV wide energy range results are given in electron volts (eV); Rydberg states, where vibrational profiles are of primary importance, are discussed in wavenumber units (cm$^{-1}$).

Our spectral study reveals detailed structure between 8.7 and 10 eV in the VUV, and dispels the suggestion[5,6,7] of 'continuous absorption' in band I. Band IV similarly described[6] as continuous, also exhibits considerable vibrational structure in our analysis below. Bands II and III of the VUV show discrete vibrational structure, previously attributed to Rydberg states.[9,10] Seccombe *et al*[9] assigned a considerable number of Rydberg states to VUV bands I to VIII; we will revise those assignments below. Later, Shastri *et al*,[11] suggested that vibrational sequences in VUV band II continue into band III, where they apparently become both separated and exhibit lower relative line widths. Our work below does not support that view. A combined VUV spectrum of **$CH_2F_2$**, shown in Fig.1 is discussed below.

Rydberg states have term values given by Equation 1:

$$\text{Term value} = \text{IE} - \text{E}_m = \frac{R}{(n-\delta)^2} \qquad (1)$$

where IE is each adiabatic ionization energy, $E_m$ is the *m*-th energy level (both in eV). *R* is the Rydberg constant (13.605 eV), *n* is the principal quantum number (PQN), while $\delta$ is the quantum defect. The lowest Rydberg states have *n* = 3; values of $\delta$ for *s*-, p-, d- and f-states are progressively smaller[4] with values $\delta \geq 0.75$, 0.4-0.6, 0.1-0.2 and 0.0-0.1 respectively. This leads to the highest term values being close to 3.1 eV for 3s Rydberg states. The PES for **$CH_2F_2$**, contains three main bands in the 12 to 21 eV region,[4] and these have been assigned to seven IE.[4] The present VUV range makes Rydberg states from several of these IE likely to occur. Ionic state vibrational patterns in the PES, often described as 'footprints', appear in many VUV spectra, and are used to identify the positions of Rydberg states in these spectra. Correlation of the PES and VUV spectra for **$CH_2F_2$**, is more complex as discussed below, and we believe these differences may be general.

These preceding studies[5-11] indicate the need for a more rigorous theoretical approach to the analysis of the VUV spectrum of **$CH_2F_2$**, as undertaken in the present paper. Our theoretical study enables us to determine the onset of several of the low-lying singlet states, by use of the time dependent density functional theory (TDDFT) and complete active space self-consistent field (CASSCF) methods. The wave-functions determined from these calculations are then used to determine the contribution of both Franck-Condon (FC) and Herzberg-Teller (HT) effects in the observed VUV bands. This type of analysis has not been performed previously for this or closely related molecules. Vertical excitation studies, where each calculated excitation energy is determined at the $X^1A_1$ ground state structure are important for this molecule, since several of the major bands in the VUV spectrum have low onset (0-0) intensities. The peak maxima here are more reliably determined than their onsets.

The decision on whether the **$CF_2$** or **$CH_2$** group is chosen for the σv axis in $C_{2v}$ symmetry, can affect the apparent excited or ionic state symmetries.[9,10,12,13] We follow our PES study[4] of **$CH_2F_2$** by using the **$CF_2$** ($y,z$) and **$CH_2$** ($x,z$) planes. The opposite choice interchanges both $B_2/B_1$ states and $b_2/b_1$ orbital symmetries. Earlier results have been reassigned to the current orientation.[4] The $X^1A_1$ ground state contains 13 doubly occupied molecular orbitals (DOMOs, 26e); however, we limit the numbering to the valence shell, and the resulting MOs are $1a_1^2$ to $4a_1^2$, $1a_2^2$, $1b_1^2$ to $2b_1^2$, $1b_2^2$ to $3b_2^2$. The highest occupied (HOMO), $2b_1^2$, is crucial in much of the Rydberg state analysis.

## II. EXPERIMENTAL AND COMPUTATIONAL PROCEDURES

### A. The VUV spectrum of $CH_2F_2$.

The sample, purchased from ABCR (99.9% catalog: AB103369), was measured using the AU-UV beam line at the ASTRID2 storage ring (Aarhus University, Denmark). The experimental setup has been described in detail previously,[3] and only a summary is included here. Monochromatic light passes through a high vacuum gas cell, and is detected using a photomultiplier tube. The intensity of light over wavelength ranges, is measured first with an evacuated cell ($I_0$) then with the sample gas present ($I_t$), and the data combined. The pressure of the gas (N) is continuously monitored using a 1 Torr heated capacitance manometer (INFICON CDG100D), with the pressure chosen to avoid over-attenuation of light. The resulting spectrum is then calculated using the well-known Beer-Lambert law, where $I_t = I_0 \exp(-Nl\sigma)$; cross-sections ($\sigma$) are given in megabarn (Mb, $10^{-22}$ m$^2$) and the path length ($l$) is 15.5 cm. The high energy toroidal grating monochromator provides photons in the range 115 to 330 nm with a resolution of 0.75 Å over the full range. The spectrum, between 3.9 eV to 10.8 eV, was recorded at room temperature with a step size varying between 0.02 to 0.5 nm, depending on the level of fine structure; no absorption was observed below 8.5 eV. A summary of the electronic states extracted from the VUV spectrum, and compared with those proposed by Seccombe et al.,[9] is shown in Table I.

**B. Simplification of the VUV spectrum by subtraction of broad structure.**

In Fig.1 (A) we combine our newly measured spectral region between 8.5 and 10.8 eV, with data by Seccombe et al..[9] This covers a wider range than that of Shastri et al.,[10] but both studies show considerable similarities. Vibrational structure, present in several regions of the current VUV spectral range, often occurs as weak structure on intense peaks, and is readily obscured. Our earlier 'subtraction' procedure, shown in Fig. 1(B), enhances these weak details.[4] Depending on the individual VUV spectral peak shape, either Gaussian functions were fitted to the region being subtracted,[4] or when a portion of the peak contained structure, such as the onset of some bands, then a Boltzmann sigmoid peak was subtracted from that region. The choice of function is completely pragmatic, in achieving the best fit to the broad structure. After numerical subtraction of these background peaks, the regular residuals were treated separately; no data is lost by the process. The residuals show sufficient cross-section count rate, for more detailed analysis to be performed. The full range 'subtraction,' shown in Fig. 1(B), shows that some very broad VUV peaks become split into more than one peak. This process enables the accurate determination of Rydberg state positions as shown in Table I. The new spectrum (Fig.

1(C) showed minor undulations in its profile between 9 and 10 eV, drawing attention to the need for subtraction. A set of vibrational peaks, far from the previous description as 'continuous absorption,[5] were obtained and are discussed below. Further detail of the subtraction process is shown in the supplementary material as SM1.

**C. Computational methods.**

Specialized aspects were performed with modules from each of the *GAUSSIAN-09 (G-09)*,[15] *MOLPRO*[16] and *GAMESS-UK*[17] suites of programs. Adiabatic excitation energies (AEE) were obtained using both TDDFT[18,19] and CASSCF in *G-09,*[15] and multi-configuration SCF (MCSCF) in *MOLPRO*.[16] Vertical excitation studies used both TDDFT, and the multi-reference multi-root doubles and singles CI method (MRD-CI)[20-22] implemented in *GAMESS-UK*.[17] Franck-Condon and Herzberg-Teller analyses were performed with the Pisa suite[23-26] internal to *G-09*.[15]

Our results were mainly performed with the optimized Coulomb fitting (second) family of basis sets known as DEF2 (default series 2);[27,28] specifically we used the DEF2-QZVP (quadruple zeta valence with polarization) and DEF2-QZVPPD (which contains additional p- and d-diffuse functions). In some calculations, we used the augmented correlation consistent quadruple zeta valence with polarization (aug-cc-pVQZ) basis sets.[29,30] In the TDDFT study we used the M11 and LC-BLYP[31-33] density functionals, which have been recommended for excited state studies.[34-37] Further detail of the theoretical methods is given in the supplementary material under SM2 through SM6.

**III. RESULTS and DISCUSSION.**

**A. The theoretical singlet excited electronic state manifolds for *$CH_2F_2$*.**

The molecular structures for the ground state equilibrium structure, and several electronically excited states for *$CH_2F_2$* using the TDDFT method are shown in Table II. Our $X^1A_1$ structure is close to the substitution structure determined by microwave spectroscopy.[38,39] Although the bond lengths show considerable variation, the bond angles, **H-C-H, F-C-F** and **H-C-F** differ from the tetrahedral value by larger amounts; there is no 'sum of angle' values in tetrahedral and related species. However, the $3^1B_1$ state is close to a regular tetrahedron. Some states shown are valence states, for example $3^1B_1$ and $1^1A_1$; this is evidenced by the high oscillator strength and widely differing structures from the Rydberg states. All AEE in Table II are relative to the energy of the $X^1A_1$ state at its equilibrium structure. In *G-09*, the excitation energy is taken as the energy difference between the $X^1A_1$ state and the state of interest, *at the equilibrium structure of the excited state*. We have corrected for this approximation. Differences between the energies of the actual ground state structure and the excited state structure, can make these corrections larger than 1 eV in some cases. For more detail, see the supplementary material as SM3.

The $^1B_1$ and $^1A_2$ manifolds, each exhibit two low-lying singlet states ($1^1B_1$ and $2^1B_1$, $1^1A_2$ and $2^1A_2$) all of which interact. The near degeneracy of the $1^1B_1$ and $1^1A_2$ states both at the TDDFT and CASSCF levels, led to configuration mixing of states of both same and differing symmetries. The resultant states ($^1B_1$, $^1A_2$ or mixed) have antisymmetric (lower energy) and symmetric (higher energy) linear combinations of configurations; for further detail, see supplementary material as SM4.

We show the potential energy surface (PESurface) for the $1^1B_1$ and $2^1B_1$ symmetric and antisymmetric states in Fig.2. The FCF angle, shows a wide difference in several electronic states in Table II, and is the structural variable most suited for these surface studies. Using the TDDFT module in *G-09* it became apparent when 'scanning' the surface (structure optimization for fixed values of this angle) led to two surfaces. The scan process uses the wave-function of the current point to project an estimate for that of the next point. This memory effect then leads to hysteresis as shown in Fig. 2 which shows the interaction of the $1^1B_1$ and $2^1B_1$ singlet states. The results for equilibrium structures in Table II are for the lower state on each surface.

However, the near degeneracy of the $1^1B_1$ and $1^1A_2$ states leads to interactions between states with different symmetry. These were studied by CASSCF in *G-09*; when applied to singlet excited states, the $X^1A_1$ state (state 1) is included in a three state-average singlet state calculation. The two open shell states (2 and 3 below) occur in linear combination, and the system is nearly degenerate, as shown:

State 2: leading configurations: +0.936 ($^1A_2$) -0.305 ($^1B_1$); Energy =9.357 eV

State 3: leading configurations: +0.936 ($^1B_1$) + 0.305 ($^1A_2$); Energy =9.414 eV

The lower energy singlet state occurs as the antisymmetric combination, which gives a close fit to a cubic surface, important to the analysis of band I of the VUV spectrum of ***CH$_2$F$_2$***. The states 2 and 3 also exhibit a curve crossing for the FCF angle of 110.5°, but distant from the equilibrium structures. Both are shown in the supplementary material in SM4.

**B. Theoretical interpretation of the overall VUV spectrum for *CH$_2$F$_2$*.**

The VUV spectrum (Fig. 3) shows the vertical excitation energies (VEE), with oscillator strengths (f(r)) superimposed; these are TDDFT calculated singlet states, determined with the aug-cc-pVQZ basis set.[29,30] The overall spectrum is dominated by $^1B_2$ and $^1B_1$ valence states, which generally have weak 0-0 bands leading to slow onsets. Numerous calculated valence states lie relatively close to the strong VUV bands IV and VII. The range of exponents in this basis set is such that the most diffuse functions will generate low-lying Rydberg ($n$ = 3) and valence states, as well as mixed states. Rydberg-valence mixing is enabled as a result. The lowest group of IEs are also shown. Rydberg states have very low f(r) (generally < 0.03), and so will not be evident in dense valence state regions of the VUV spectrum as in Figs. 1 and 3.

However, the presence of seven IE in the region shown, shows that a high density of Rydberg states must contribute to the spectral substructure. A description of the vertical Rydberg states, determined by the MRD-CI method is shown in the supplementary material under SM7.

Several 0-0 bands for low-lying states, together with their oscillator strengths, are shown in Fig. 4. The nine harmonic frequencies for each state are shown in the supplementary material under SM8; those participating in the Franck-Condon and Herzberg-Teller analyses are shown in Tables below.

**C. Individual bands of the VUV spectrum.**

Although band I in the observed spectrum, centred on 74860 cm$^{-1}$ (9.281 eV) has previously been described as continuous absorption,[5] the 'subtracted' spectrum, in Fig. 5 shows a complex but irregularly spaced set of peaks. This is a plot of the regular residuals after removal of a Gaussian 'best fit' function from the main VUV band I. The subtracted region shown, between 9.00 and 9.60 eV, contains 950 data points, proving that noise is not important. Several peak separations alternate in magnitude with values near 440 and 350 cm$^{-1}$. The mean separations of 390 cm$^{-1}$, are much smaller than the lowest TDDFT calculated mode 4 ($a_1$) of the $1^1B_1$ state (563 cm$^{-1}$), and even smaller than the CASSCF mixed state, where mode 4 is 677 cm$^{-1}$.

Band I, previously assigned to a Rydberg state,[5,6,8,9] clearly does not present the expected IE$_1$ PES footprint.[4] The multiplet of 14 maxima in the $X^2B_1$ state,[4] mean separation 1069 cm$^{-1}$, would be readily observed if present in band I of the VUV spectrum. The PES will fit under band I, after a lowering of its energy by 3.948 eV; this PES shift to superimpose the VUV spectrum, yields a potential Rydberg state having $n = 3$ with $\delta = 1.14$, in agreement with earlier spectral work.[9] However, this must be an incomplete analysis, since the lack of similarity in the spectral profiles cannot be ignored. Further, a more realistic Rydberg state assignment using $n = 3$, is with the higher band II. Our earlier PES assignment[4] used the sophisticated MP4SDQ method, which we regard as rigorous; thus the distinction between band I and the PES is clear.

The VUV region below 16 eV has been analyzed in terms of numerous electronic states. We start our analysis with a detailed study of Band II. This has an apparent similarity to the lowest band observed in the PES; this was attributed to the $^2B_1$ state. That assignment is consistent with assignment of the VUV band II as a Rydberg state. We will find that the PES in these Rydberg states, is only one factor in the VUV band profiles, and is the reason why we discuss VUV band I last.

**D. Band II of the VUV spectrum.**

This sharp group of vibrational peaks contrasts with the band I structure, and gives a reasonably good fit to the PES $X^2B_1$ envelope, as shown in Fig.6. This ionic state has an 0-0 band at 12.746 eV (102805 cm$^{-1}$). We assign this to a Rydberg state as $n = 3$ with $\delta = 0.818$, after a term value shift of the PES by 2.860 eV. This is consistent with a 3s-Rydberg state.[46,47] This band was previously assigned by Seccombe *et al*[9] as a 3p-Rydberg state; we believe that $\delta = 0.818$ is too high for that assignment in both our own experience, and Robin.[46,47] Most p-Rydberg states have $\delta$ in the range 0.3 to 0.6. Our assignment with the choice of $n = 3$ is valid, since the preceding band I is largely valence in character, and other factors (below) are important. The calculated AEE (9.886 eV, 79554 cm$^{-1}$), is slightly lower than the apparent onset at 10.027 eV, while the VUV intensity is maximal at 10.285 eV (82954 cm$^{-1}$).

Our initial vibrational analysis for VUV band II as a Rydberg state, follows directly from our earlier PES study of the $X^2B_1$ state,[4] (shown in Fig. 3 of Reference 4). The chosen half-width at half-maximum (HWHM, 75 cm$^{-1}$) for the calculated ionic state in Fig.6, is a typical value for a PES band which is well separated from its neighbours.[4] The broadening of the main vibrational bands with increasing energy, is a result of the number of vibrational modes enclosed in each peak. The principal vibrations, summarised in Reference 4, are dominated by contributions from ($a_1$) modes 1, 2 and 3, and their combinations. These can be expressed generally as $1^a 2^b 3^c$ where a, b, c are quanta. The lowest $a_1$ vibration, mode 4, is much less evident. The relative narrowing of the peaks between 10.9 and 11.0 eV, part of the next band system is notable.

The analysis above is insufficient, since there is weak structure on the low energy side of each peak, and evidence of splitting of the main peaks. This points to inadequacy of the routine assumption that a pure Rydberg state, should show a VUV spectral profile identical with the ionic state. We have directly analysed the $2^1B_1$ singlet state using the TDDFT method, and this is completely independent of comparison with the PES. This analysis as a singlet state allows additional factors to be considered. We have evaluated the individual Franck-Condon (FC), Herzberg-Teller (HT), and combined (FCHT) contributions for the $2^1B_1$ state. The combined FCHT theoretical spectrum is shown in Fig. 7. Further, the centre of gravity of the separate FC and HT terms are different, with most HT at the higher energy end. Generally, the FC manifold is lower in intensity by a factor of over 30, increasing to over 100 for the higher bands. The main intensity is determined by the HT rather than FC terms. The leading terms shown in Table III, exhibit a much more complex situation; there are many more vibrational states for $2^1B_1$ at the FCHT level, than are required to describe the PES band. Further, the number of quanta excited is much higher, with many vibrational states having 4 or more quanta. Although almost all the FC states in Table III also occur in the HT manifold, the reverse is not true. None of the non-symmetric modes are stimulated at the vibronic fundamental level.

Although the absorption and fluorescence spectra of VUV band I show some similarity,[9] there is a fundamental difference between 9.8 and 11 eV, where the fluorescence intensity is close to zero. Our FCHT calculations show that the fluorescence spectrum is almost entirely limited to the $X^1A_1$ state modes $3a_1$ and $2a_1$ for each of the $2^1B_1$ and $1^1A_1$ states in this energy range. This resolves the experimental difference between absorption and fluorescence, and *inter alia* supports the current assignment for the VUV spectral band II. The lowest calculated $^1B_2$ state is predicted to lie in the gap between bands II and III, but is not evident, owing to a weak onset.

**E. Band III of the VUV spectrum.**

This group of vibrational peaks are fundamentally different from those of band II, both from increased sharpness, uneven spacing and irregular intensity, as shown in Fig.8. The marked narrowing of the vibrational bands in the onset to band III at 10.912 eV (88017 cm$^{-1}$) was shown in Fig. 6. It corresponds to a full width at half maximum (FWHM) of 117 cm$^{-1}$ in contrast to the lower peaks of band II which have FWHM 260 cm$^{-1}$. Since the assignment of band II is fundamentally a classical Rydberg state exhibiting the $X^2B_1$ state profile, an alternative assignment is necessary for band III.

Seccombe *et al.*[9] assigned it to a Rydberg state based on IE$_2$ (this is assigned to the $A^2B_2$ state in our $C_{2v}$ axis orientation of ***CH₂F₂***).[4] The A ionic state, is a set of weak peaks on a rising background, which is part of three overlapping ionic states in the PES.[4] The spacing of vibrations for this ionic state vary across the band;[4] direct measurement of the peak separations gives a value of 542 cm$^{-1}$ for the central region, declining to 505 cm$^{-1}$ at the end of the sequence at 15.237 eV. The present set of (nine) peaks in the VUV range of 88271 to 91505 cm$^{-1}$, shows a lower mean separation of 404 cm$^{-1}$, with standard deviation 29 cm$^{-1}$. This clearly does not fit the vibrational structure of the $A^2B_2$ state in the PES of ***CH₂F₂***.

In our present analysis of this VUV region, almost every peak can be accounted for by three super-positions of the $X^2B_1$ PES, with 0-0 bands at 10.913, 10.962 and 11.017 eV, as is shown in Fig. 8. The 0-0 bands give term values 1.833 (n = 3, δ = 0.273), 1.784 (n = 3, δ = 0.237) and 1.729 eV (n = 3, δ = 0.194) respectively in Fig.8; this leads to the assignment in this region, as two 3p- and (possibly) one 3d-Rydberg state, all based on the $X^2B_1$ ionic core. One 3p state would have $^1A_2$ symmetry, and is optically forbidden. Relatively weak sharp doublets, occur near 11.263 and 11.394 eV as well as elsewhere, and are not accounted for; these may indicate further, very weak, states under this portion of the VUV envelope.

The separation of band III into three parts, now contains the $3^1A_1$ state as lowest member. The FCHT analysis of this state using the TDDFT procedure, is shown in Table IV. Although this shows low levels of HT participation for the lowest lying vibrations in the manifold, the HT proportion rises with increasing energy from below 1% to above 6%. Indeed, the separate FC and HT manifolds, in Fig. 9, show that each main FC vibrational band has a lower energy onset of HT nature. This has the effect of broadening the overall vibrational profiles, and the onset subsidiary peaks mentioned above. The combined FC + HT profiles are too complex to overlay on the VUV spectrum. The HT contributions cannot be ignored however, since the number of HT vibrations excited, is very much greater than those for the FC model, and the cumulative effect is considerable. For this state, the totals are 2640 (FC) and 82087 (HT). Non-symmetric modes 6 and 7 are stimulated in the HT, even close to the 0-0 band. The importance of HT in our results for other states are similar. A calculated valence state occurs at 11.518 eV with high oscillator strength (0.102). It is the $3^1B_1$ state, assigned to the observed peak at 11.732 eV (94633 cm$^{-1}$). Although this will account for some absorption peaks in band III, it cannot cover the complete range, and is not an alternative to the above triple analysis.

**F. Band IV of the VUV spectrum.**

After subtraction of the under-lying broad peak in this VUV range, and processing the regular residuals, two sets of peaks with maxima at 11.690 and 12.397 eV (94300 and 100000 cm$^{-1}$ respectively) are obtained, as seen in Fig. 10. Clearly, each set of peaks could be offset by one or more quanta. We assume these are Rydberg states, but since these are too close to the AIE of the $X^2B_1$ ionic state, they will be associated with the $A^2B_2$ ionic state. The slow onset and wide range, fulfils this requirement, and leads to the assignment as $3b_23s$ ($n = 3$, with $\delta = 0.895$) and $2b_23p$ ($n = 3$, with $\delta = 0.595$) states respectively. Seccombe et al.[9] observed one of these states, which they ascribed to a $3b_23p$ Rydberg, but the value of $\delta = 0.8$ is too high for a 3p-state. These fits show residual structure, apparent in Fig. 10, suggesting the appearance of valence states, or Rydberg states based on higher AIE may also be present.

The calculated FC profile consists of a sequence in mode 4 ($a_1$) with frequency 485 cm$^{-1}$, where the central maximum occurs after seven quanta (Table V). Subsidiary frequencies are largely (binary and tertiary) combination bands of modes 1 and 4. The contribution of HT intensity is relatively small, but several non-symmetric modes are activated.

### G. Bands V and VI in the VUV spectrum.

These are poorly defined, but enhanced by subtraction of a sloping baseline; both are twin peaks, as seen in Fig. 1(B). Our extracted maxima (Table I) are significantly different from those by Seccombe et al.[9] who assigned them to Rydberg states involving IE$_3$ ($B^2A_1$). That PES state is not well defined,[4] but probably lies close to the calculated value of 15.218 eV. Its lowest Rydberg state $4a_13s$ (assuming $n = 3$, with $\delta = 0.8$) would lie close to 12.4 eV, the region required here. Seccombe et al.[9] assigned VUV band V as the $4a_13s$ Rydberg state, but gave an unacceptably low value of $\delta = 0.48$; their proposed alternative was $1a_23s$, but this is optically forbidden and also incorrect. The appearance of bands V and VI is consistent with numerous overlapping Rydberg state vibrations. However, insufficient resolution precludes a detailed analysis.

### H. Band VII in the VUV spectrum.

This is the most intense band in the VUV spectrum, stretching from 14.5 to 16.7 eV, with a broad maximum near 15.7 eV. The subtraction process indicates overlap with VUV band VI, as is evident in Fig. 1 (B). The presence of one or more valence and Rydberg states under the envelope seems certain; even a high density of Rydberg states is unlikely to generate such a high cross-section. In addition, underlying individual components of band VII in Fig. 1(B), show varied vibrational structure; the individual maximum near 15.14 eV shows substantial vibrational structure, while the peak near 16.17 eV is apparently unstructured.

Subtraction of a Boltzmann function from the leading edge of band VII, leaves a well-defined set of maxima shown in Fig. 11. The slow rise of the untreated peak, makes the 0-0 band poorly defined in the subtracted version. The probable onset is near 14.527 eV (117172 cm$^{-1}$). The group of 11 peaks, are a single vibrational sequence, with a mean separation of 529 cm$^{-1}$ (median 533; standard deviation 19 cm$^{-1}$). This energy is too high for an $n$s-Rydberg state based on any of IE$_1$ to IE$_4$ of the PES (X, A, B, C states), which occur in the PES range below 15-16 eV.[4] It follows that if this structure is a Rydberg state, it must arise from the third band of the PES lying between 18 and 20 eV, which contains the D$^2$B$_2$, E$_2$A$_1$ and F$^2$A$_2$ states.[4]

In our PES analysis,[4] both the onset and the trailing edges on band III were subjected to subtraction of appropriate Boltzmann sigmoid peaks, in order to enhance weak vibrational structure. The resulting appearances[4] are very similar to that in Fig. 11. The slow onset is consistent[4] with either D$^2$B$_2$ or F$^2$B$_1$ ionic states, shown in Figures 7 to 9 of reference 4;[4] the E$^2$A$_1$ state, predicted to have a relatively high 0-0 band, can be omitted on that basis.

The onset of PES band III,[4] like Fig. 11, gives a mean peak separation, 691 cm$^{-1}$ (median 682 cm$^{-1}$). Similar treatment of the trailing edge of PES band III, gives regular residuals with two vibration frequencies, 623 cm$^{-1}$ (median 614 cm$^{-1}$), and a weaker sequence, with separation 235 cm$^{-1}$, which becomes submerged in the main sequence as the energy increases. Thus, there is a fundamental difference in the spacing between both sets of vibrations in the PES and band VII of the VUV spectrum. We believe that these frequency differences exclude ionic states in bands

I, II or III of the PES as being responsible for a Rydberg state with vibration frequency 529 cm$^{-1}$ in VUV band VII. Our calculated a$_1$ modes for the ionic states are: 841 and 433 (2$^2$B$_2$), 845 and 361 (2$^2$A$_1$) and 1015 and 455 (2$^2$B$_1$);[4] none of these are close to the values for VUV band VII. Overall, we conclude that the Fig. 11 vibrations are valence in character.

Two $^1$A$_1$ valence states have high oscillator strengths, and are in the band VII energy range. The lower at 16.552 eV with f(r) 0.3097 has a$_1$ frequencies 583 and 1075 cm$^{-1}$; the higher at 17.153 eV with f(r) 0.397 has a$_1$ frequencies 498 and 1318 cm$^{-1}$. Although these are relatively close to the Fig. 11 value, and may provide an interpretation of it, both have high 0-0 bands, which does not appear to be the case with Fig. 11. More appropriate are $^1$B$_2$ states, owing to their low onset intensity. The calculated $^1$B$_2$ state, occurring at 13.537 eV, with f(r) 0.1425 is typical. Mode 9 (b$_2$) occurs at 558 cm$^{-1}$, and also occurs in multiple combination bands in the FCHT analysis. This is an attractive assignment for band VII in the VUV spectrum.

A general point emerges in these FCHT analyses, in relation to the frequency separation between the 4a$_1$ and 3a$_1$ modes of the various singlet states. The harmonic frequencies shown in the supplementary material as SM8 (as Table SM8), show that modes 6 to 9 (b$_1$ + b$_2$) can occur in the canonical sequence between the 4a$_1$ and 3a$_1$ modes. Herzberg-Teller contributions appear to offer an alternative solution to some observed vibrational levels which are not readily interpreted in higher energy regions such as band VII.

The next IE of **CH$_2$F$_2$**$^+$ above the three PES bands I, II and III, is the G$^2$A$_1$ state (IE 24.0 eV).[48,49] Although poorly described, it appears to show vibrational structure. This would be consistent with the 3t$_2$$^{-1}$ PES bands of both **CH$_4$** and **CF$_4$** near 22.0 eV, both of which show well-defined vibrational splitting. A 10-member sequence for the latter[14] has interval 645 cm$^{-1}$. If this proposition concerning the G$^2$A$_1$ state is correct, Rydberg states with a similar vibrational sequence, would appear in the VUV spectrum, but beyond the energy range of the present experiment.

**I. Assignment of band I of the VUV as a single state.**

The FCHT simulation of band I, using the equilibrium structure of the $1^1B_1$ state TDDFT wavefunction, generates a very complex envelope of many transitions, as shown in Fig. 12. Vibrational analysis, shows both Franck-Condon (FC) and Herzberg-Teller (HT) contributions occur. A summary, in Table VI, of the lowest set of cold band vibrations with their intensities, shows that the most intense set of vibrations involve combinations of all the $a_1$ modes (1 to 4). Most vibrations appear in both FC and HT sequences. The calculated profile shows that decrease of the vibrational intensity on the high energy tail of the calculated band is slow, leading to asymmetry, which also occurs on the observed VUV band I. The calculated FC profile for the $1^1B_1$ state is dominated by modes 3, 2 and 1 (783, 1289 and 2198 cm$^{-1}$), while the only single overtone excited in the HT profile is mode 6 ($b_1$, 1892 cm$^{-1}$). Although the FC modes are dominant for the vibrational states close to the 0-0 band, the HT modes become more important in the tail, as was found with band II; most vibrational states are common to FC and HT manifolds.

In summary, if band I of the VUV spectrum can be considered as a single state, then it is a very complex mixture in which all $a_1$ fundamentals, and binary, tertiary and quartic combination bands, together with HT contributions participate. However, this single state analysis does not account for the observed vibrations in Fig.5. The band I profile obtained by superimposing Gaussian functions on the stick values in Fig. 12 to the VUV spectrum, only becomes nearly smooth, for a FWHM of 675 cm$^{-1}$. This value is much larger than might be expected for a single state, but is consistent with highly overlapping states. For example, our studies of *PhI*,[40] showed that FWHM of 300 and 400 cm$^{-1}$ were required for the $A^2A_2$ and $B^2B_2$ overlapping states respectively. Kwon et al.[41] observed that a bandwidth of 400 cm$^{-1}$, occurred for the origin of the $B^2B_2$ state of *PhI* in their MATI spectrum. We believe that a similar situation may occur here, since we have a near degenerate second ($^1A_2$) forbidden state, with an anticipated major overlap of their vibrational wave-trains.

**J. Alternative theoretical assignment of band I of the VUV.**

Formally, $^1A_2$ singlet states are widely ignored in VUV spectroscopy, except where vibronic coupling leads to an allowed transition. Here we consider the effects of mixing of states which exhibit almost exact degeneracy, but are of different symmetries; specifically $^1B_1$ with the $^1A_2$ singlet state manifold. Study of this type of interaction is routine in *G-09*, using 'state-average' CASSCF calculations.[15] The $X^1A_1$ state is included (as root 1) in a three-root state-average analysis, while the two open shell states (2 and 3) were described in Section IIIA above. Further description is given in the supplementary material under SM9. The CASSCF vibration frequencies under these circumstances, shown in Table VII, differ considerably from the TDDFT values, especially with modes 3 ($a_1$) and 9 ($b_2$). These do not give an acceptable interpretation of the vibrational structure in Fig.5 for band I. Such an analysis must await a jet-cooled VUV or related study.

**IV. CONCLUSIONS**.

Our narrow range VUV study of ***CH₂F₂***, at the highest resolution yet performed for this compound, shows that bands I and IV are clearly not 'continuous absorption' as previously described. We have combined the spectrum with a dataset, kindly made available to us by Seccombe *et al*.,[9] so that we can perform a more detailed analysis of the wider spectral range. The spectrum by Shastri *et al*.[10] covers part of the combined range with Seccombe *et al*.,[9] but shows no additional features.

The subtraction of broad structure has enabled us to process the regular residuals in unprecedented detail, and has relatively enhanced weak fine structure across the spectrum. We believe that many previously reported VUV spectra may also disclose fine structure if treated in this manner; such data is potentially recoverable and available for analysis. The importance of vertical studies is clear; the VUV spectrum of **CH$_2$F$_2$** is dominated by $^1B_1$ and $^1B_2$ states, where the 0-0 bands are very weak. The actual onset is uncertain in some cases, and the apparent 0-0 band could be offset by one or more vibrational quanta. We believe that our current analysis of the VUV spectrum for **CH$_2$F$_2$** provides a model for interpretation of many VUV spectra.

Our recent high-resolution PES study together with its vibrational analysis,[4] was not available to previous VUV studies.[5-11] This has now led to detailed analysis of vibrational structure in several VUV bands having Rydberg state character. We have been able to identify several Rydberg states by overlay of PES spectral bands, suitably shifted in energy, to coincide with the VUV state. The assignments provide a more acceptable set of quantum defects relative to previous studies,[9,10] as shown in Table I. The Rydberg part of the VUV spectrum is dominated the $X^1B_1$ and $A^2B_2$ states identified in the PES. The most intense VUV band VII, in the 15 to 17 eV region, is primarily valence in character; fine structure on its low energy side does not contain an ionic state footprint. Two previous sets of Rydberg state assignments,[9,10] identified Rydberg state energies by comparison with early determinations of IE, using vertical values from the PES where necessary. Several of these earlier reported quantum defects are outside normal ranges.[46,47] Another potential assignment, $1a_23s$,[9] is not optically allowed and is also in error. Shastri et al.,[10,11] assigned all vibrational peaks in the range 10 to 11.5 eV to three long progressions. We believe this is incorrect; none of the large energy range, irregular spacing or the widely differing spectral line widths are accounted for in that manner.

Subtle differences between PES and VUV have been found, by direct calculation of the Rydberg singlet state vibrational profile, where we use both Franck-Condon and Herzberg-Teller procedures. Currently PES analyses contain FC but not HT terms. These differences should be regarded as the norm rather than the exception for low PQN Rydberg states, and we expect such differences will occur widely.

In contrast to the earlier studies,[5,9,10] we find that the VUV onset between 8.75 and 10 eV, described as band I, can be deconvoluted by our subtraction process, to a complex set of broad peaks. Each of these may be composites of several vibrations. Bands I (and IV) are clearly not 'continuous absorption'.[5,9] We believe that the unusual nature of absorption band I may result from a singlet state interaction of the $1^1B_1$ state with the $1^1A_2$ state; an example of a second-order Jahn-Teller interaction[52,53,54] without spin-orbit coupling.[55] The near degeneracy of the individual $1^1A_2$ and $1^1B_1$ states, leads to closely coupled energy surfaces. There is loss of distinction between allowed and forbidden states in the linear combinations of these two states, which occur as symmetric and antisymmetric combinations. However, direct theoretical interpretation of the deconvoluted VUV band I cannot be made at this time. We agree with early studies,[5] that the high density of states in band I certainly causes spectral congestion, and this may contribute to the lack of resolution of band I.

Potential energy surfaces, are widely generated by automated scanning procedures, using an appropriate variable (here the FCF angle). The direction of scan across the variable range can be important. The 'scan' procedure projects the wave-function for the current point, to become an estimate for the next point; this leads to a memory effect and hysteresis in potential energy surfaces. The adiabatic excitation energies for several singlet states determined by TDDFT methods exhibit this aspect. In the current study, two curves occur for symmetric and antisymmetric combinations of $1^1B_1$ and $2^1B_1$ states, and these need to be scanned separately. This effect should be general for molecules with symmetry. There is a danger that this proposition, when applied to earlier potential energy surfaces, may show them to be non-unique, or not even the lowest in energy.

It is essential to correct the apparent adiabatic excitation energies given for TDDFT and CASSCF in *G-09*, to values based on the separate $X^1A_1$ equilibrium state structure, as is normal in spectroscopy. The *G-09* approach determines the AEE from the energy for both ground and excited states at the same geometry. These corrections can easily be of the order of 1 eV.

The AEE have been analyzed by a combination of Franck-Condon and Herzberg-Teller methods. The low-lying band II is largely a mixture of FC and HT vibrations, and this is unexpected since the footprint of the PES $X^2B_1$ is present; previously this PES band was interpreted by solely FC methods. In all VUV bands studied, both FC and HT types occur alongside each other, in varying proportions. Even where the HT components are overall relatively small, several vibrational peaks are based on both $b_1$ and $b_2$ vibrations. The vibrational patterns have been compared with our recent PES profiles, previously analyzed[4] in considerable detail. Although this allows assignment to be made for several Rydberg states, the direct comparison of singlet state vibrational structure with the PES is only an approximation, since these ionic state analyses are purely based on FC contributions.

When referring to VUV spectra generally, we predict that many electronic states, which may have been rejected as Rydberg states owing to differences from the PES profile, may be wrongly classified. The two types of spectra must not be expected to be identical for all Rydberg states, although that assumption may often be correct. We have already shown that attempts to correlate PES profiles with single vibrations in the ground state, are almost certain to be inaccurate for polyatomic molecules. This is equally true for singlet states in the VUV region. Suitable theoretical methods were unavailable for many early assignments of PES and VUV spectra. This will necessitate re-analysis of many spectra in future years.

Our most successful basis sets are from the DEF2-QZVPPD series;[27,28] the high levels of polarization included, are important to achieve realistic vibration frequencies.

**SUPPLEMENTARY MATERIAL**

**SM1. The VUV 'subtraction' process.**

**SM2. Computational methods expanded. Basis sets.**

**SM3. Molecular orbital interactions in the excited states of *$CH_2F_2$*.**

**SM4. Effect of symmetry and near degeneracy on the excited states.**

**SM5. Position of Rydberg state functions.**

**SM6. Conical intersections and avoided crossings.**

**SM7. Singlet Rydberg states for *$CH_2F_2$* using the MRD-CI method**

**SM8. The harmonic frequencies for the singlet states studied.**

**SM9. The lowest singlet state of the 3-root CASSCF state-average calculation.**

**SM10. Supplementary Material References.**

**ACKNOWLEDGEMENTS**

We thank the following: Professor R. P. Tuckett (Birmingham,UK),for a copy of his dataset of the VUV absorption spectrum; Professor Malgorzata Biczysko for helpful discussions; Edinburgh Parallel Computing Centre for access to Gaussian-09 on their computer, cirrus.epcc.ed.ac.uk. Plotting and numerical analysis used the OriginLab Corporation, Northampton, MA01060, USA, suite (Origin2016, 64bit, Sr2).

**Table I. The present VUV spectral analysis (adiabatic (A) and vertical (V)) compared with that of Seccombe *et al*.[9]**

| | Present assignment | | | | | Seccombe *et al*[9] | | | |
|---|---|---|---|---|---|---|---|---|---|
| Band | Energy (A or V) / eV | State | PQN $n$ | Quantum defect δ | Band | Energy (A or V) / eV | State | PQN $n$ | Quantum defect δ |
| I | 8.767 (A) 9.281(V) | $2b_1a^*$ | (3) | (1.143)[a] | I | 9.28 | $2b_13s$ | 3 | 1.16 |
| II | 9.734(A) 10.385(V) | $2b_13s$ | 3 | 0.818 | II | 10.01 | $2b_13p$ | 3 | 0.76 |
| III | 10.913(A), 10.962(A), 11.017(A) | $2b_13p$ $2b_13p$ $2b_13d$ | 3, 3, 3, | 0.273 0.237 0.194 | III | 10.91 | $3b_23s$ | 3 | 1.08 |
| IV | 11.158 (A),11.690(V) 11.255(A), 12.397(V) | $2b_23s$ $2b_23p$ | 3 3 | 0.895 0.595 | IV | 12.38 | $3b_23p$ | | 0.82 |
| V | 13.20(A),13.69(A) 13.41(V),13.85(V) | | | | V | 13.56 | $4a_13s$ | 3 | |
| VI | 14.76 (A); 15.12 (V) | | | | VI | 13.86 | $4a_13p$ | 3 | 0.29 |
| VII | 15.75(A) 16.17(V) | $3b_23s$ $3b_23s$ | 3 | 1.04 | VII | 15.72 | $1b_13s$ | 3 | 1.00 |

**Table II.** Adiabatic excitation energies (eV) and oscillator strengths f(r) for the singlet states, using the DEF2-QZVPPD and DEF2-QZVP basis sets for Rydberg and valence states respectively. The apparent TDDFT excitation energies in *G-09*, have been corrected to AEE by using the energy difference between the excited state and the $X^1A_1$ state at their respective equilibrium structures. The $X^2B_1$ ionic state results use the DEF2-QZVP basis set.

| Energy / eV | f(r) | Symmetry[a] | Leading configurations | H-C | C-F | HCH | FCF | HCF |
|---|---|---|---|---|---|---|---|---|
| - | - | $X^1A_1$ | SCF | 1.079 | 1.332 | 108.9 | 108.3 | 113.0 |
| - | - | $X^2B_1$ | $2b_1^1$ | 1.177 | 1.263 | 85.3 | 116.3 | 112.8 |
| 8.6284 | 0.0 | $1^1A_2$ | $2b_1b_2^*$ | 1.133 | 1.406 | 91.8 | 96.5 | 118.0 |
| 8.7111 | 0.0704 | $1^1B_1$ | $2b_1a_1^*$ | 1.197 | 1.271 | 74.6 | 115.3 | 115.2 |
| 9.7338 | 0.0004 | $2^1B_1$ | $2b_1a_1^*$ | 1.179 | 1.295 | 75.5 | 122.3 | 112.4 |
| 10.2209 | 0.0725 | $1^1A_1$ | $2b_1b_1^*$ | 1.173 | 1.269 | 94.3 | 115.8 | 111.2 |
| 10.7044 | 0.0903 | $2^1A_1$ | $2b_1b_1^*$ | 1.179 | 1.272 | 95.2 | 115.4 | 111.1 |
| 10.8103 | 0.0329 | $1^1B_2$ | $3b_2a_1^*$ | 1.094 | 1.388 | 126.5 | 86.6 | 109.1 |
| 11.0541 | 0.0996 | $3^1A_1$ | $2b_1b_1^*$ | 1.179 | 1.272 | 95.2 | 115.4 | 111.1 |
| 11.5180 | 0.1021 | $3^1B_1$ | $2b_1a_1^*$ | 1.164 | 1.291 | 115.2 | 114.2 | 106.9 |
| 11.7357 | 0.0 | $2^1A_2$ | $2b_1b_2^*$ | 1.153 | 1.275 | 88.3 | 114.1 | 113.0 |
| 11.8224 | 0.0015 | $4^1B_1$ | $2b_1a_1{*}$ | 1.162 | 1.270 | 84.2 | 118.3 | 112.4 |
| 11.9624 | 0.0244 | $1^1B_2$ | $6a_15b_2{*}$ | 1.073 | 1.721 | 135.0 | 83.8 | 106.5 |
| 11.9645 | 0.0123 | $4^1A_1$ | $4a_1a_1{*}$ | 1.109 | 1.423 | 124.0 | 119.8 | 103.6 |
| 12.2986 | 0.0190 | $4^1A_1$ | $6a_1a_1{*}$ | 1.140 | 1.425 | 104.6 | 118.3 | 121.2 |
| 12.3738 | 0.0082 | $5^1A_1$ | $6a_1a_1{*}$ | 1.114 | 1.424 | 123.3 | 119.6 | 103.8 |
| 12.4244 | 0.0149 | $2^1B_2$ | $6a_16b_2{*}$ | 1.072 | 1.822 | 156.0 | 88.6 | 89.5 |
| 13.5370 | 0.1425 | $3^1B_2$ | $2b_1a_2{*}$ | 1.085 | 1.421 | 121.5 | 90.2 | 110.2 |
| 16.4714 | 0.2084 | $4^1B_2$ | $1a_24b_1{*}$ | 1.083 | 1.440 | 115.9 | 99.7 | 110.0 |
| 16.5521 | 0.3097 | $6^1A_1$ | $6a_1a_1{*}$ | 1.103 | 1.422 | 120.6 | 112.5 | 106.0 |
| 17.1527 | 0.3977 | $7^1A_1$ | $2b_1b_1{*}$ | 1.097 | 1.453 | 113.7 | 111.8 | 107.8 |
| 18.4303 | 0.1928 | $5^1B_2$ | $1a_24b_1{*}$ | 1.097 | 1.465 | 111.0 | 111.0 | 111.0 |
| 18.8582 | 0.3038 | $8^1A_1$ | $4b_2b_2{*}$ | 1.202 | 1.489 | 105.4 | 99.0 | 113.2 |
| 20.7099 | 0.1832 | $6^1B_2$ | $3b_2a_1{*}$ | 1.077 | 1.585 | 120.6 | 89.9 | 110.5 |
| 22.6780 | 0.7372 | $7^1B_2$ | $2b_1a_2{*}$ | 1.157 | 1.342 | 110.7 | 106.7 | 107.3 |

**Table III. The $2^1B_1$ state, showing the lowest group of vibrations with quanta excited, occurring in the Franck-Condon and Herzberg-Teller analyses. The mode sequence numbers are 1-4 ($a_1$), 5($a_2$), 6-7($b_1$), 8-9 ($b_2$).**

| Vibration and quanta | Frequency / cm$^{-1}$ | Intensity | | |
|---|---|---|---|---|
| | | Franck-Condon | Herzberg-Teller | HT/FC |
| 0 | 0 | 1293 | - | -- |
| $4^1$ | 263 | 3 | - | -- |
| $3^1$ | 1022 | 11 | 365 | 33 |
| $2^1$ | 1128 | 9 | 449 | 50 |
| $3^14^1$ | 1284 | 14 | 448 | 32 |
| $3^2$ | 2043 | 21 | 757 | 36 |
| $2^2$ | 2257 | 9 | 707 | 79 |
| $3^24^1$ | 2306 | 27 | 924 | 34 |
| $1^1$ | 2585 | 3 | - | -- |
| $3^3$ | 3065 | 26 | 918 | 35 |
| $3^34^1$ | 3328 | 21 | 741 | 35 |
| $2^3$ | 3385 | 6 | 623 | 104 |
| $3^4$ | 4087 | 14 | 548 | 39 |
| $3^44^1$ | 4349 | 20 | 760 | 38 |
| $2^4$ | 4513 | - | 339 | -- |
| $3^5$ | 5108 | 9 | 392 | 44 |
| $3^54^1$ | 5371 | 14 | 542 | 39 |
| $2^23^34^1$ | 5584 | 49 | 4397 | 90 |
| $3^6$ | 6130 | 5 | - | -- |
| $1^12^23^2$ | 6885 | 45 | 4792 | 106 |
| $1^12^13^34^1$ | 7041 | 64 | 5000 | 78 |
| $1^12^23^24^1$ | 7148 | 53 | 5561 | 105 |

| Vibration and quanta | Frequency / cm$^{-1}$ | Intensity | | |
|---|---|---|---|---|
| | | Franck-Condon | Herzberg-Teller | 100HT/FC |
| $1^1 2^2 3^3$ | 7907 | 51 | 5685 | 111 |
| $1^1 2^2 3^3 4^1$ | 8169 | 60 | 6554 | 109 |

**Table IV.** The $3^1 A_1$ state, showing the lowest group of vibrations with quanta excited, occurring in the Franck-Condon and Herzberg-Teller analyses. The mode sequence numbers are 1-4 (a$_1$), 5(a$_2$), 6-7(b$_1$), 8-9 (b$_2$).

| Vibration and quanta | Frequency / cm$^{-1}$ | Intensity | | |
|---|---|---|---|---|
| | | Franck-Condon | Herzberg-Teller | 100HT/FC |
| 0 | 0 | 28570 | 38 | 0 |
| $4^1$ | 607 | 1117 | 0 | 0 |
| $7^1$ | 893 | | 122 | -- |
| $3^1$ | 1036 | 25750 | 81 | 0 |
| $2^1$ | 1286 | 33760 | 37 | 0 |
| $1^1$ | 1400 | 14660 | 123 | 1 |
| $7^2$ | 1787 | 575 | 0 | 0 |
| $3^1 7^1$ | 1929 | | 121 | -- |
| $3^2$ | 2071 | 6044 | 55 | 1 |
| $6^1$ | 2086 | | 85 | -- |
| $2^1 7^1$ | 2180 | | 188 | -- |
| $2^2$ | 2572 | 31540 | 17 | 0 |
| $7^3$ | 2680 | | 8 | -- |
| $2^1 3^1 7^1$ | 2787 | | 181 | -- |
| $3^3$ | 3107 | 614 | 13 | 2 |
| $6^1 2^1$ | 3372 | | 131 | -- |
| $1^1 3^1$ | 3436 | 22980 | 194 | 1 |
| $2^2 7^1$ | 3466 | | 154 | -- |
| $1^1 2^1$ | 3686 | 28930 | 159 | 1 |
| $2^3$ | 3859 | 19830 | 0 | 0 |
| $1^2$ | 4800 | 4231 | 100 | 2 |
| $2^4$ | 5145 | 9392 | 0 | 0 |
| $1^2 3^1$ | 5836 | 7625 | 148 | 2 |
| $1^2 2^1$ | 6086 | 9135 | 151 | 2 |
| $2^5$ | 6431 | 2600 | 0 | 0 |
| $1^3$ | 7200 | 472 | 32 | 7 |
| $2^6$ | 7717 | 1005 | 0 | 0 |

**Table V.** The $1^1 B_2$ state, showing a selected group of vibrations with quanta excited, occurring in the Franck-Condon or Herzberg-Teller analyses. The mode sequence numbers are 1-4 (a$_1$), 5(a$_2$), 6-7(b$_1$), 8-9 (b$_2$). The low 0-0 band and slow onset are typical of the $^1 B_2$ states.

| Vibration and quanta | Frequency / cm$^{-1}$ | Intensity | | |
|---|---|---|---|---|
| | | Franck-Condon | Herzberg-Teller | 100HT/FC |
| $4^2$ | 970 | 712 | | |
| $4^3$ | 1454 | 1579 | 12 | 1 |
| $3^1 4^1$ | 1575 | 179 | | -- |
| $4^4$ | 1939 | 4170 | 39 | 1 |
| $3^1 4^2$ | 2060 | 1004 | 12 | 1 |

| Vibration and quanta | Frequency / cm⁻¹ | | | |
|---|---|---|---|---|
| $4^5$ | 2424 | 7594 | 59 | 1 |
| $3^1 4^3$ | 2545 | 3229 | 45 | 1 |
| $4^6$ | 2908 | 8881 | 126 | 1 |
| $5^1 4^4$ | 3001 | | 107 | -- |
| $3^1 4^4$ | 3030 | 5630 | 107 | 2 |
| $3^2 4^2$ | 3150 | 580 | | -- |
| $4^7$ | 3393 | 12060 | 192 | 2 |
| $5^1 4^5$ | 3485 | | 149 | -- |
| $3^1 4^5$ | 3514 | 11360 | 149 | 1 |
| $3^2 4^3$ | 3635 | 2066 | 0 | 0 |
| $4^8$ | 3879 | 9185 | 191 | 2 |
| $3^1 4^6$ | 3999 | 11480 | 245 | 2 |
| $3^2 4^4$ | 4120 | 3114 | | -- |
| $3^3 4^2$ | 4240 | 288 | | -- |
| $4^9$ | 4363 | 10110 | 228 | 2 |
| $3^1 4^7$ | 4455 | | 219 | -- |
| $3^1 4^7$ | 4484 | 17270 | 210 | 1 |
| $3^2 4^5$ | 4605 | 6961 | | -- |
| $3^3 4^3$ | 4725 | 701 | | -- |
| $4^{10}$ | 4848 | 8573 | 160 | 2 |
| $5^1 4^8$ | 4939 | | 277 | -- |
| $3^1 4^8$ | 4969 | 19460 | | -- |
| $3^2 4^6$ | 5089 | 11110 | | -- |
| $4^{11}$ | 5333 | 5009 | 157 | 3 |
| $5^1 4^9$ | 5424 | | 172 | -- |
| $3^2 4^7$ | 5574 | 10140 | | -- |
| $4^{12}$ | 5818 | 3680 | 81 | 2 |
| $3^2 4^8$ | 6059 | 12670 | | -- |
| $4^{13}$ | 6303 | 1574 | 68 | 4 |
| $4^{14}$ | 6788 | 1029 | 45 | 4 |
| $4^{15}$ | 7272 | 538 | 20 | 4 |
| $4^{16}$ | 7757 | 195 | 11 | 6 |

**Table VI. The $1^1 B_1$ state, showing the lowest group of vibrations with quanta excited, occurring in the Franck-Condon and Herzberg-Teller analyses. The mode sequence numbers are 1-4 ($a_1$), 5($a_2$), 6-7($b_1$), 8-9 ($b_2$).**

| Vibration and quanta | Frequency / cm⁻¹ | Intensity | | |
|---|---|---|---|---|
| | | Franck-Condon | Herzberg-Teller | 100HT/FC |
| 0 | 0 | 1707 | -- | -- |
| $3^1$ | 783 | 4073 | 73 | 2 |
| $2^1$ | 1289 | 4239 | 80 | 2 |
| $3^1; 4^1$ | 1345 | 616 | -- | -- |
| $3^2$ | 1566 | 6238 | 120 | 2 |
| $2^1 4^1$ | 1851 | 702 | -- | -- |
| $6^1$ | 1892 | -- | 117 | Pure HT |
| $2^1 3^1$ | 2072 | 13580 | 272 | 2 |
| $3^2 4^1$ | 2128 | 494 | -- | -- |
| $1^1$ | 2198 | 1338 | 67 | 5 |
| $3^3$ | 2349 | 4963 | 104 | 2 |
| $2^2$ | 2578 | 5226 | 109 | 2 |
| $2^1 3^1 4^1$ | 2634 | 1624 | -- | -- |
| $6^1 3^1$ | 2675 | -- | 409 | Pure HT |
| $2^1 3^2$ | 2855 | 12750 | 275 | 2 |
| $1^1 3^1$ | 2980 | 4977 | 253 | 5 |
| $3^4$ | 3132 | 2076 | -- | -- |
| $2^2 1^1$ | 3141 | 924 | -- | -- |
| $6^1 2^1$ | 3181 | -- | 270 | Pure HT |
| $2^2 3^1$ | 3361 | 16270 | 362 | 2 |
| $2^2 3^2$ | 4144 | 20450 | 491 | 2 |
| $1^1 2^1 3^1$ | 4270 | 12440 | 641 | 5 |
| $2^3 3^1$ | 4650 | 12890 | 317 | 2 |
| $1^1 2^1 3^2$ | 5053 | 18500 | 981 | 5 |
| $2^3 3^2$ | 5433 | 15560 | 413 | 3 |
| $1^1 2^2 3^1$ | 5559 | 15310 | 806 | 5 |
| $1^1 2^1 3^3$ | 5835 | 14220 | 781 | 5 |
| $1^1 2^2 2^2$ | 6342 | 21930 | 1191 | 5 |
| $1^1 2^3 3^1$ | 6848 | 12360 | 668 | 5 |
| $1^1 2^2 3^3$ | 7125 | 15970 | 903 | 6 |
| $1^1 2^3 3^2$ | 7631 | 17040 | 953 | 6 |
| $1^1 2^4 3^1$ | 8137 | 12960 | 721 | 6 |
| $1^1 2^3 3^3$ | 8414 | 11730 | 686 | 6 |
| $1^2 2^2 3^2$ | 8539 | 13350 | 1471 | 11 |
| $1^2 2^2 3^3$ | 9322 | 10070 | 1112 | 11 |
| $1^2 2^3 3^2$ | 9828 | 12300 | 1322 | 11 |

*Table VII. The CASSCF harmonic frequencies (cm$^{-1}$) for the $1^1B_1$ state coupled with the $1^1A_2$ state, compared with the corresponding TDDFT single excitation CI values in $C_{2V}$. The CASSCF values are generally larger. Modes 3 ($a_1$) and 9 ($b_2$) differ considerably between the two methods, and are nearly degenerate in the CASSCF calculation. If these are omitted, the other frequencies have TDDFT = 0.903CASSCF – 46 cm$^{-1}$, with adjacent $R^2$ 0.993*

| Method | Modes | 1 | 2 | 3 | 4 | 5 | 6 | 7 | 8 | 9 |
|---|---|---|---|---|---|---|---|---|---|---|
| CASSCF | Symmetry | $a_1$ | $a_1$ | $a_1$ | $a_1$ | $a_2$ | $b_1$ | $b_1$ | $b_2$ | $b_2$ |
| | Frequency | 2430 | 1464 | 1213 | 677 | 1138 | 2223 | 924 | 1703 | 1219 |
| TDDFT | Frequency | 2198 | 1289 | 787 | 563 | 1037 | 1892 | 747 | 1486 | 855 |

*Figure Captions*

*Figure 1. The vacuum ultraviolet spectrum of $CH_2F_2$; A: Digital data kindly provided by R. P. Tuckett.[9] B: The subtracted spectrum (in blue) is discussed below. C: The new low energy region (in red). Subtraction of the underlying broad structure in C and the residual spectrum, is discussed below.*

*Figure 2. The potential energy surfaces for the pair of $1^1B_1$ and $2^1B_1$ singlet states in symmetric (symm) and antisymmetric (asymm) linear combinations. This scanning procedure leads to a hysteresis effect over the two surfaces. A similar picture also occurs in the $^1A_2$ manifold.*

*Figure 3. The VUV spectrum of $CH_2F_2$ with the TDDFT VEE and oscillator strengths, with adiabatic ionization energies marked as vertical blue lines. The TDDFT, showing 200 lowest singlet states, were determined using the aug-cc-pVQZ basis. This basis contains some relatively diffuse functions, so that low-lying Rydberg states and states of mixed valence and Rydberg character are included. Valence states show very variable intensity, but generally much higher f(r) than Rydberg states. Hence valence states dominate the energy range shown. In contrast, 'pure' Rydberg states exhibit generally very low f(r) and would have little impact on the valence state results shown. $^1A_2$ states (f(r) = 0) are also shown.*

*Figure 4. The VUV spectrum with low-lying adiabatic excitation energies, using the DEF2-QZVPPD basis set within the TDDFT method, are superimposed as a stick diagram (in red).*

*Figure 5. The onset (Band I) of absorption by $CH_2F_2$ where the unprocessed VUV spectrum is shown in black. A baseline ramp was used to flatten the spectrum, and a 'best fit' Gaussian function was then subtracted from the data. This removes the broad background, and the Band previously described[5] as 'continuous absorption' is now seen (in red) as a complex set of vibrational states.*

*Figure 6. The VUV spectrum of the 10.0 to 11.0 eV region; the new spectrum is shown in blue, with the Seccombe et al in black. The experimental profile of the $X^2B_1$ state is super-imposed in red, indicating the presence of a Rydberg state, but evidence of further structure is clear, both from peak maxima and onset of each peak.*

*Figure 7. The calculated $2^1B_1$ state covers the slightly wider range of 9.5 to 11.25 eV of the VUV spectrum. The $2^1B_1$ state (shown in red) is made up of both Franck-Condon and Herzberg-Teller contributions, and hence more complex than the $1^2B_1$ ionic state. The earlier PES analysis was limited to FC contributions. All intensities in this Paper are given as the molar absorption coefficient (in $dm^3.mol^{-1}.cm^{-1}$). The blue curve is the stick diagram convoluted with a Gaussian profile of HWHM = 75 $cm^{-1}$.*

*Figure 8. The VUV spectrum of the 10.8 to 11.8 eV region. The profile of the $X^2B_1$ state is superimposed three times (red, green and blue). These define 3 Rydberg states, with 0-0 bands at 10.913, 10.962 and 11.017 eV respectively, and account for almost all the peaks in the spectral region shown. The VUV peaks are sharper than those of the PES experiment, possibly a result of differing lifetimes. Evidence of further states is apparent from the pairs of vibrations at 90800 and 92922 $cm^{-1}$.*

*Figure 9. The calculated profile of the Franck-Condon (red) and much weaker Herzberg-Teller (blue) contributions to the $3^1A_1$ state.*

*Figure 10. Band IV of the VUV spectrum after subtraction of the underlying broad peak; adjacent averaging of the data points has been performed to reduce noise in the spectrum. Two Rydberg states are disclosed by incorporating the PES profiles (in red and blue).*

*Figure 11. The onset to Band VII after subtraction of a Boltzmann sigmoid curve to remove broad structure. The apparent 0-0 band is indicated, but might be lower in energy by one or more vibrational quanta.*

*Figure 12. The FCHT simulation (in red) of Band I absorption by $CH_2F_2$ assuming a single electronic state. This spectrum requires a FWHM of approximately 600 cm$^{-1}$ to generate the observed VUV profile; the value shown (in blue) has FWHM 500 cm$^{-1}$. A strong interaction with another state, now identified as the $1^1A_2$ state, which is effectively degenerate with $1^1B_1$ may lead to an interpretation of the observed vibrational profile.*

# Palmer et al Fig. 1

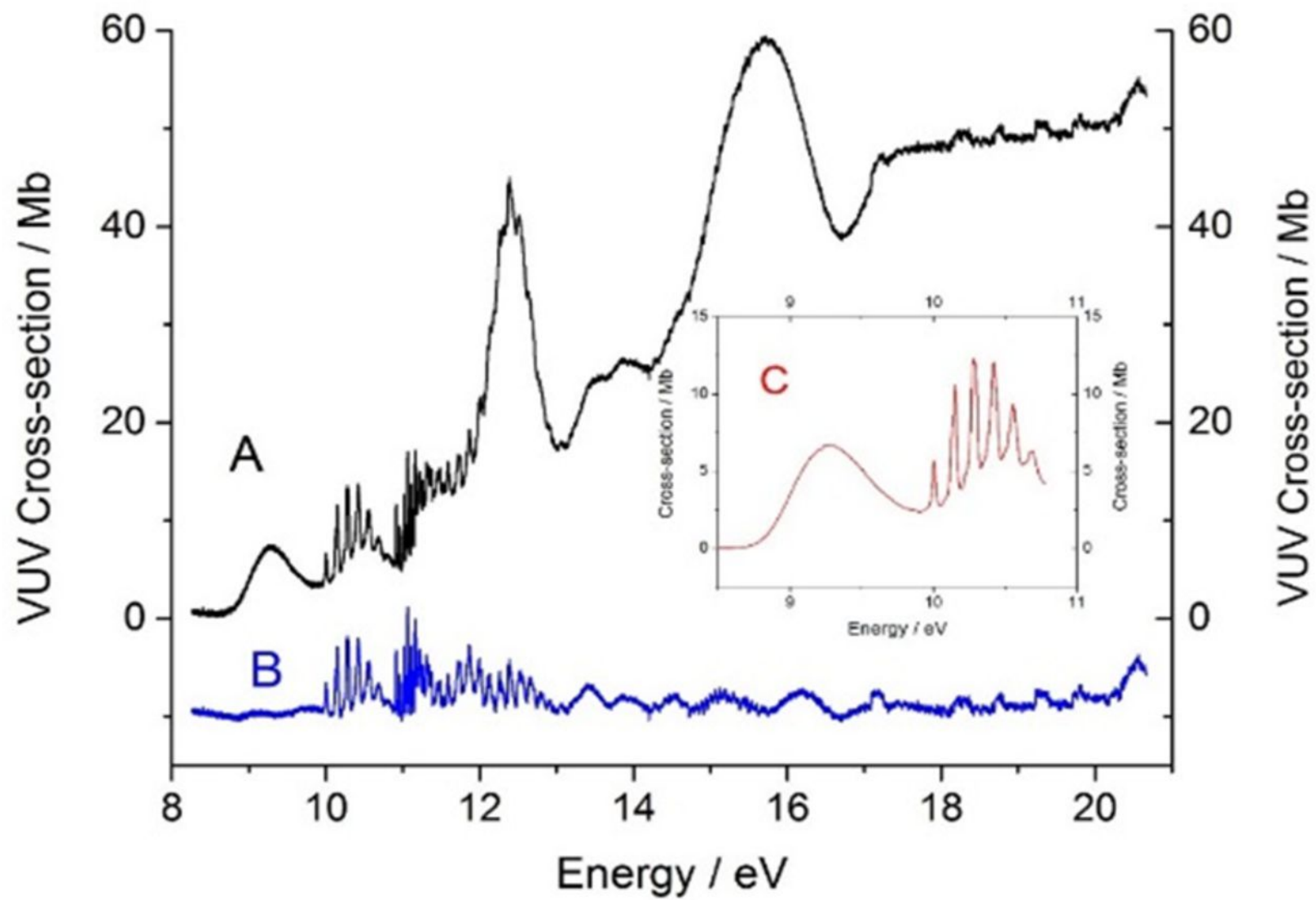

# Palmer et al Fig. 2

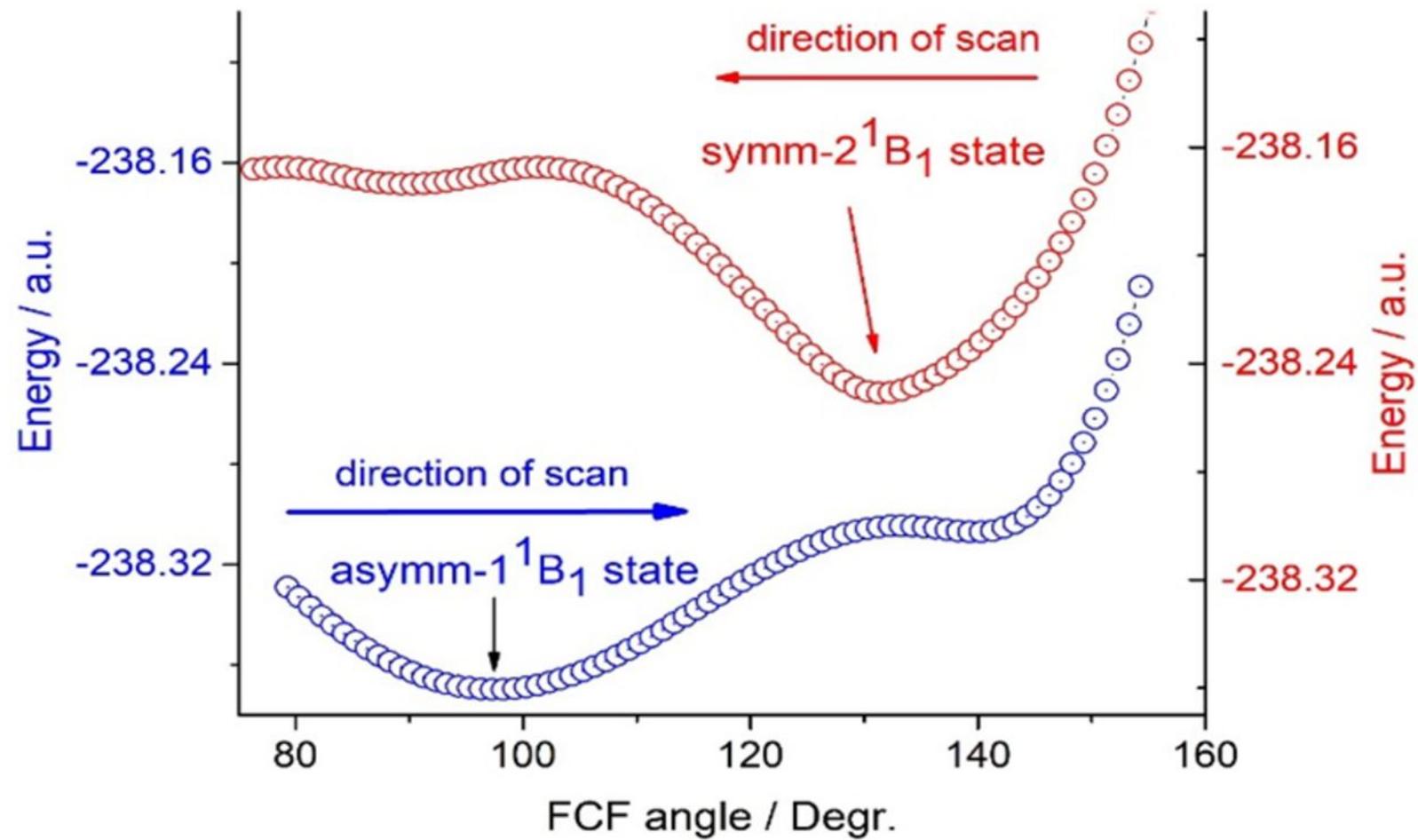

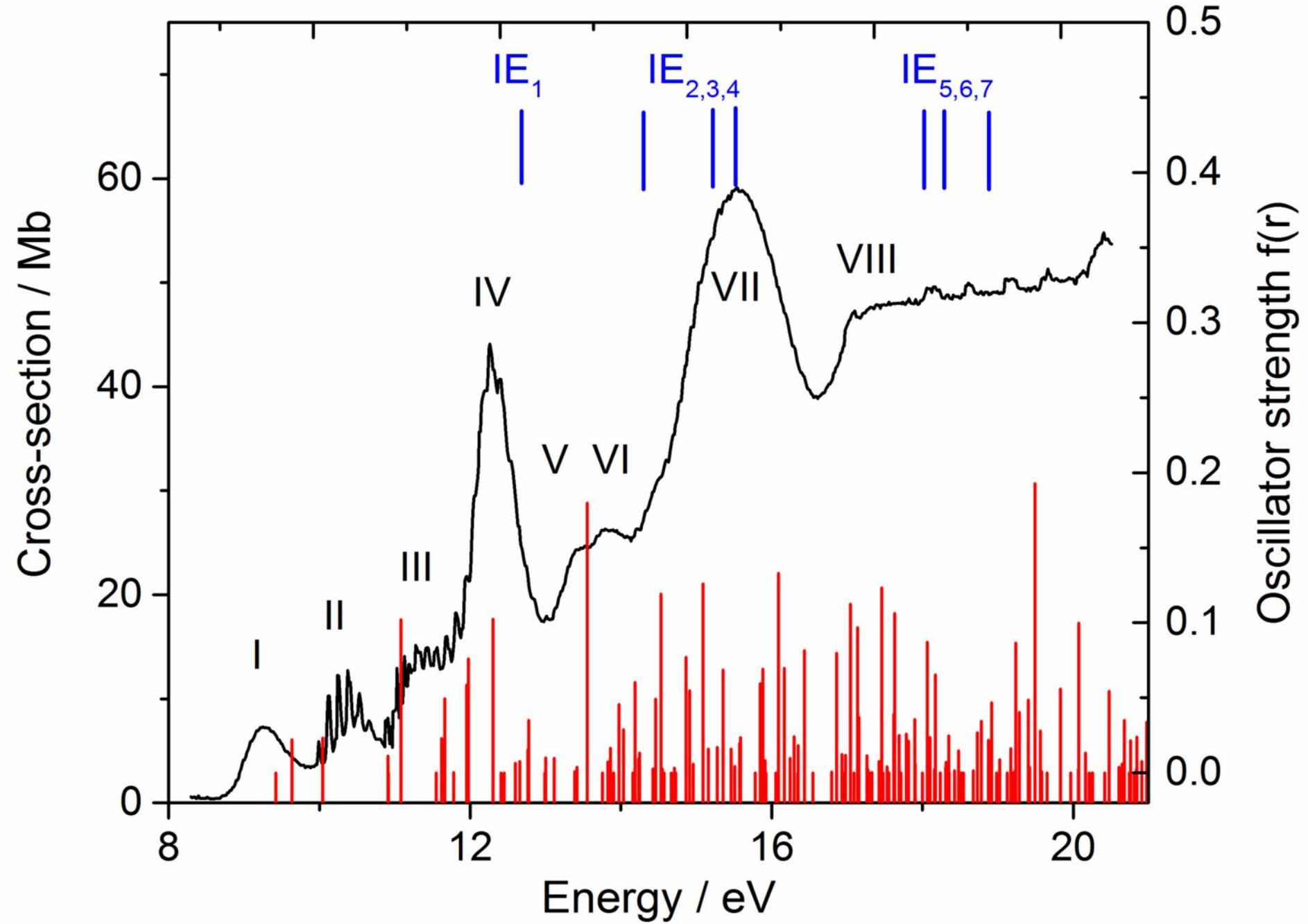

Palmer et al Fig.3

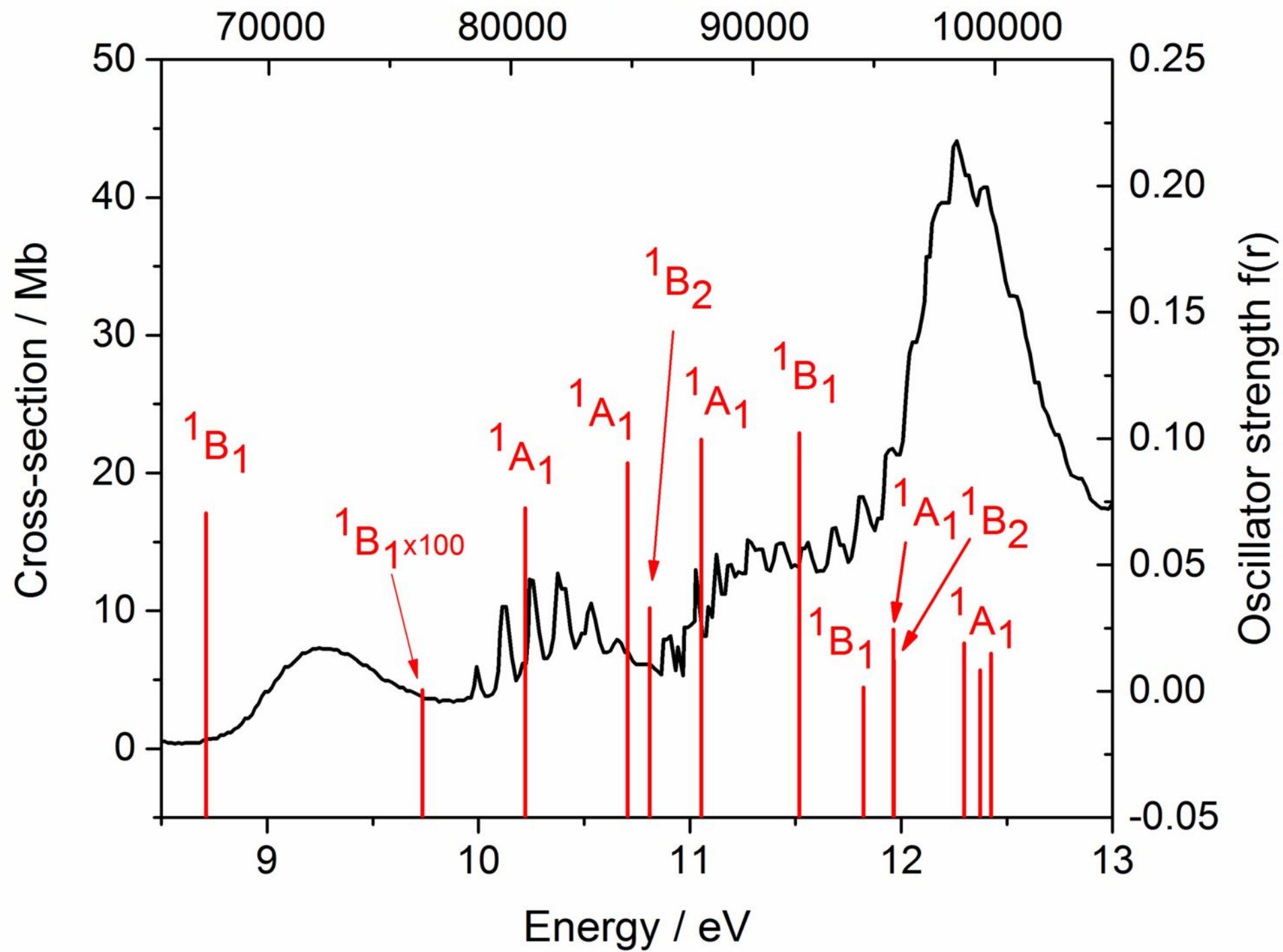

Palmer et al Fig. 4

# Palmer et al Fig. 5

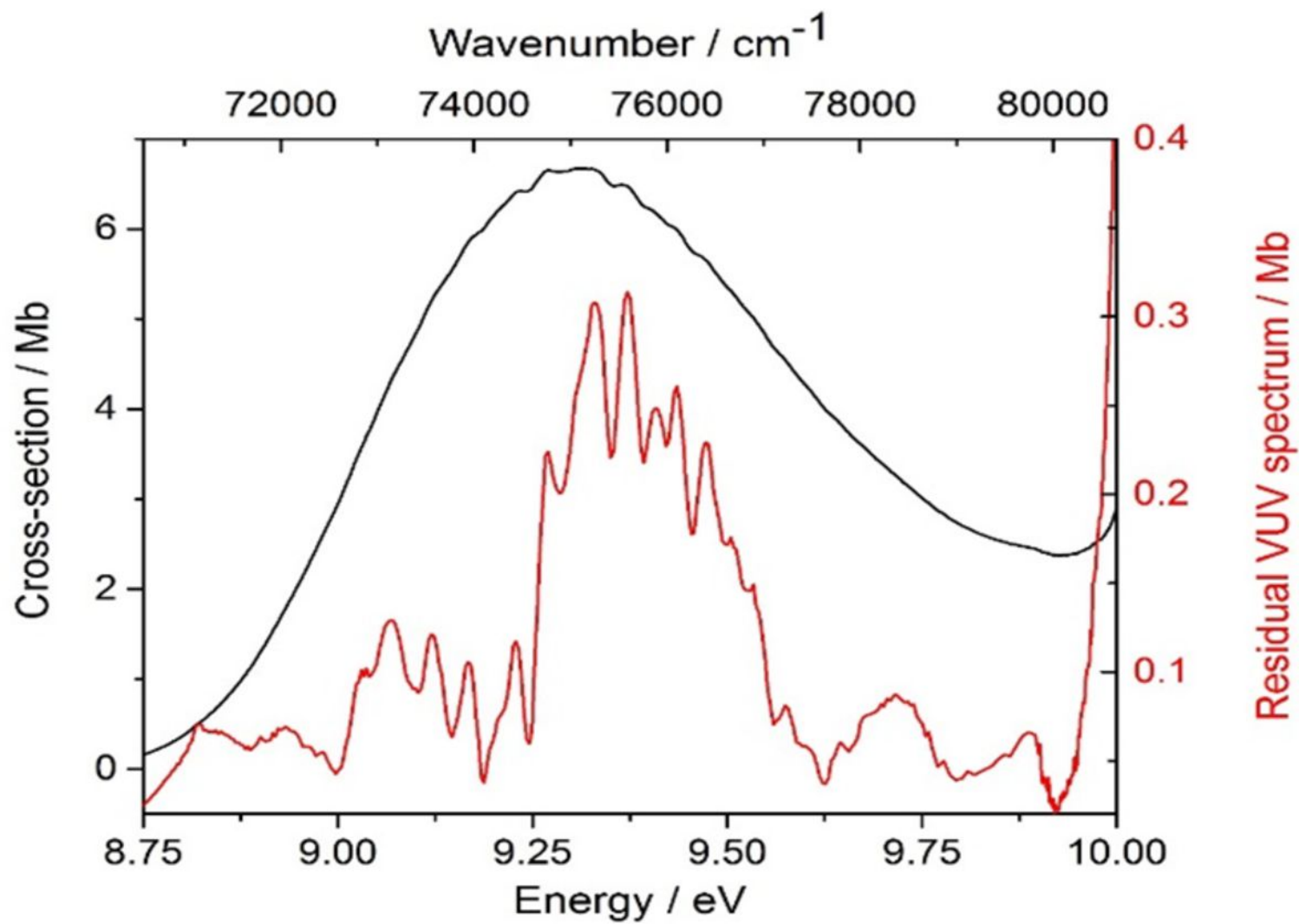

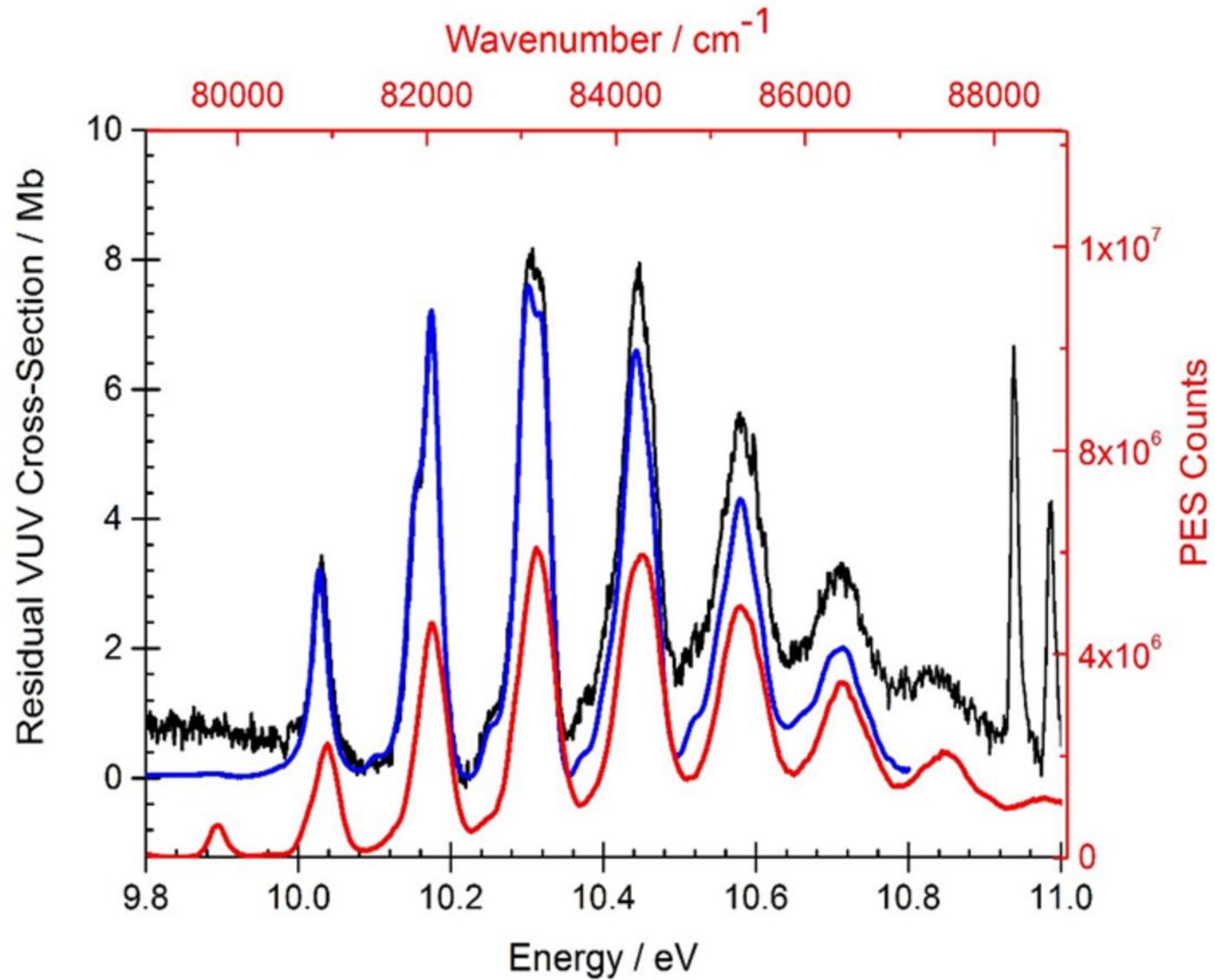

# Palmer et al Fig.7

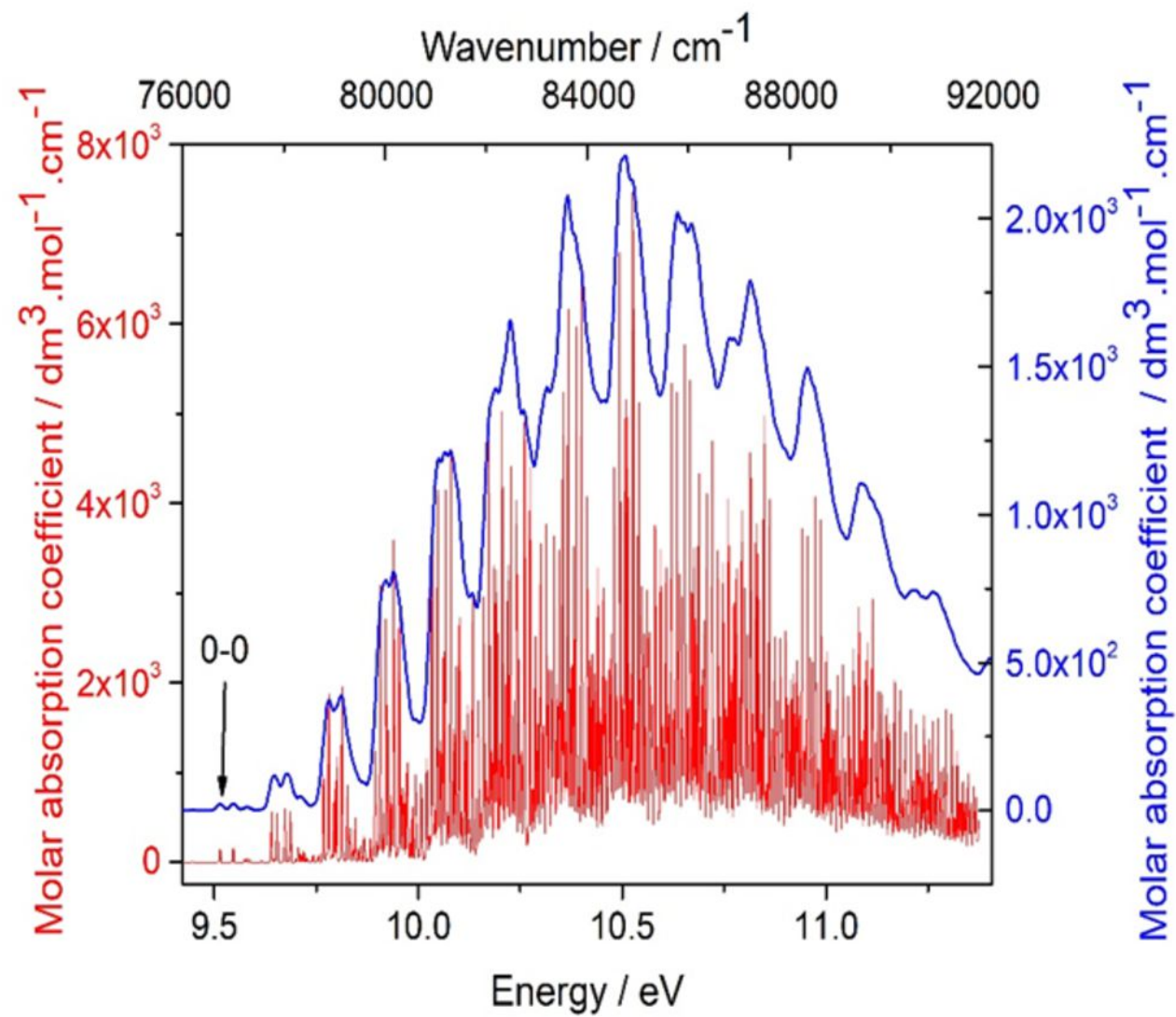

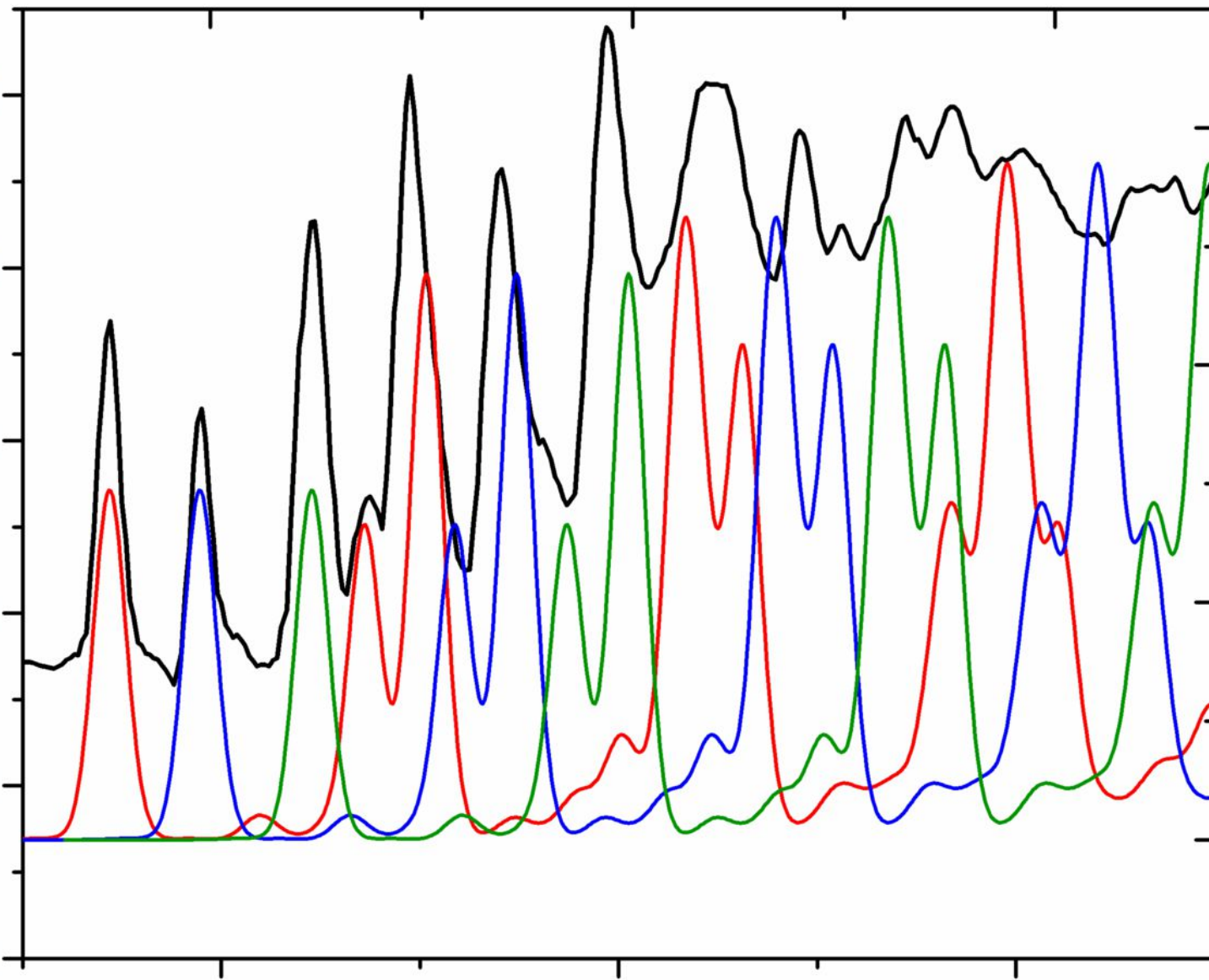

Palmer et al Fig.8

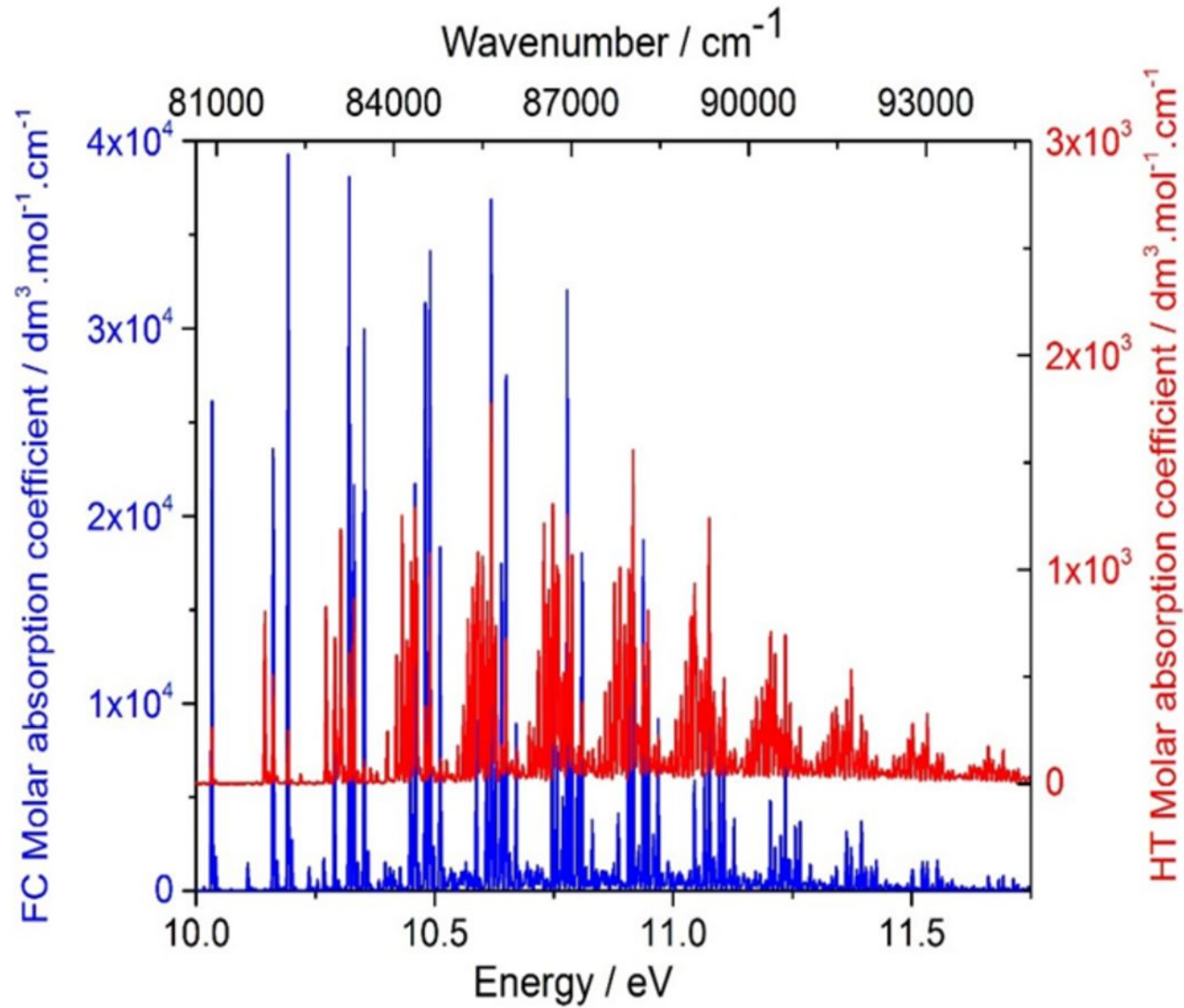

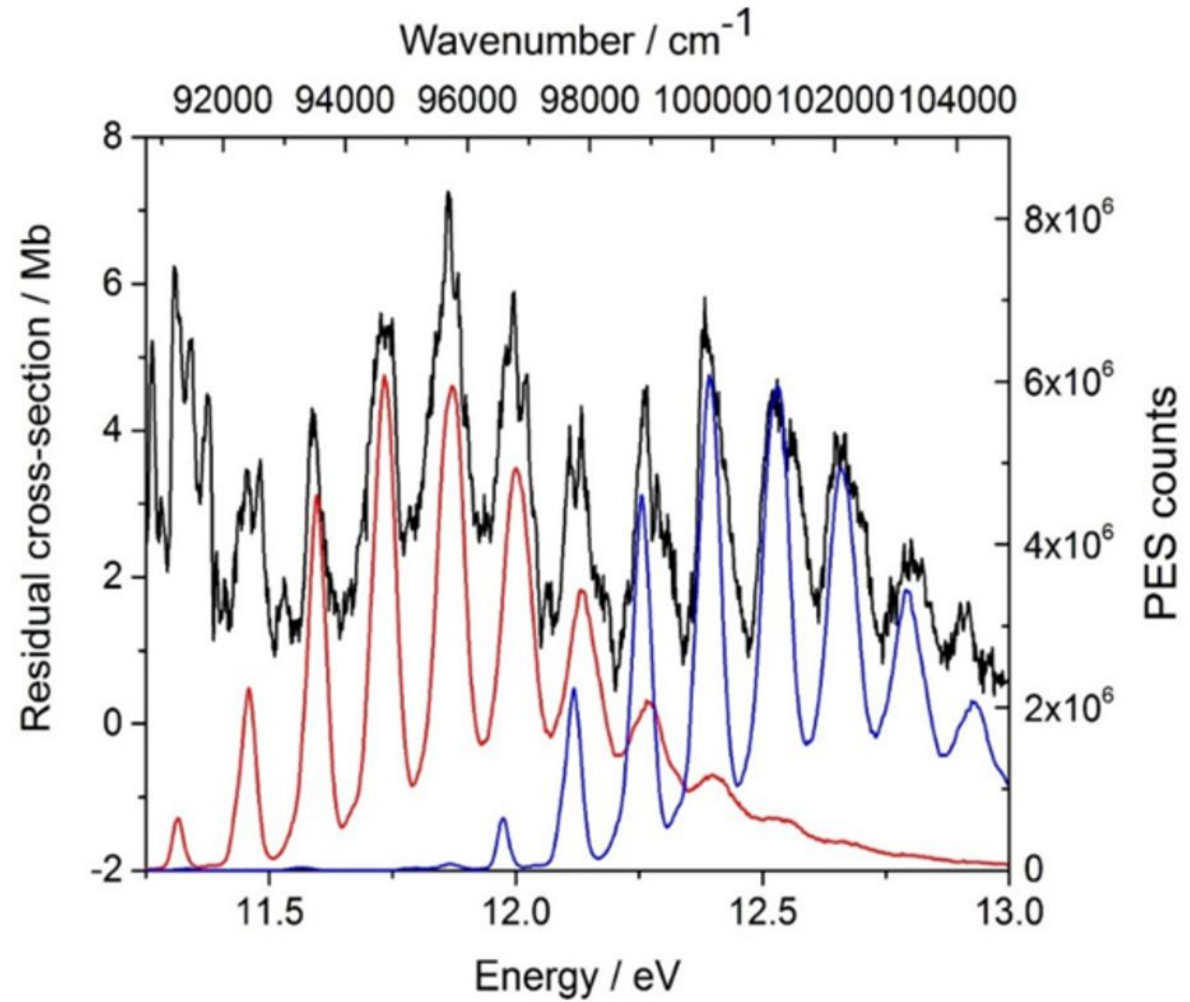

Palmer et al Fig.10

# Palmer et al Fig.11

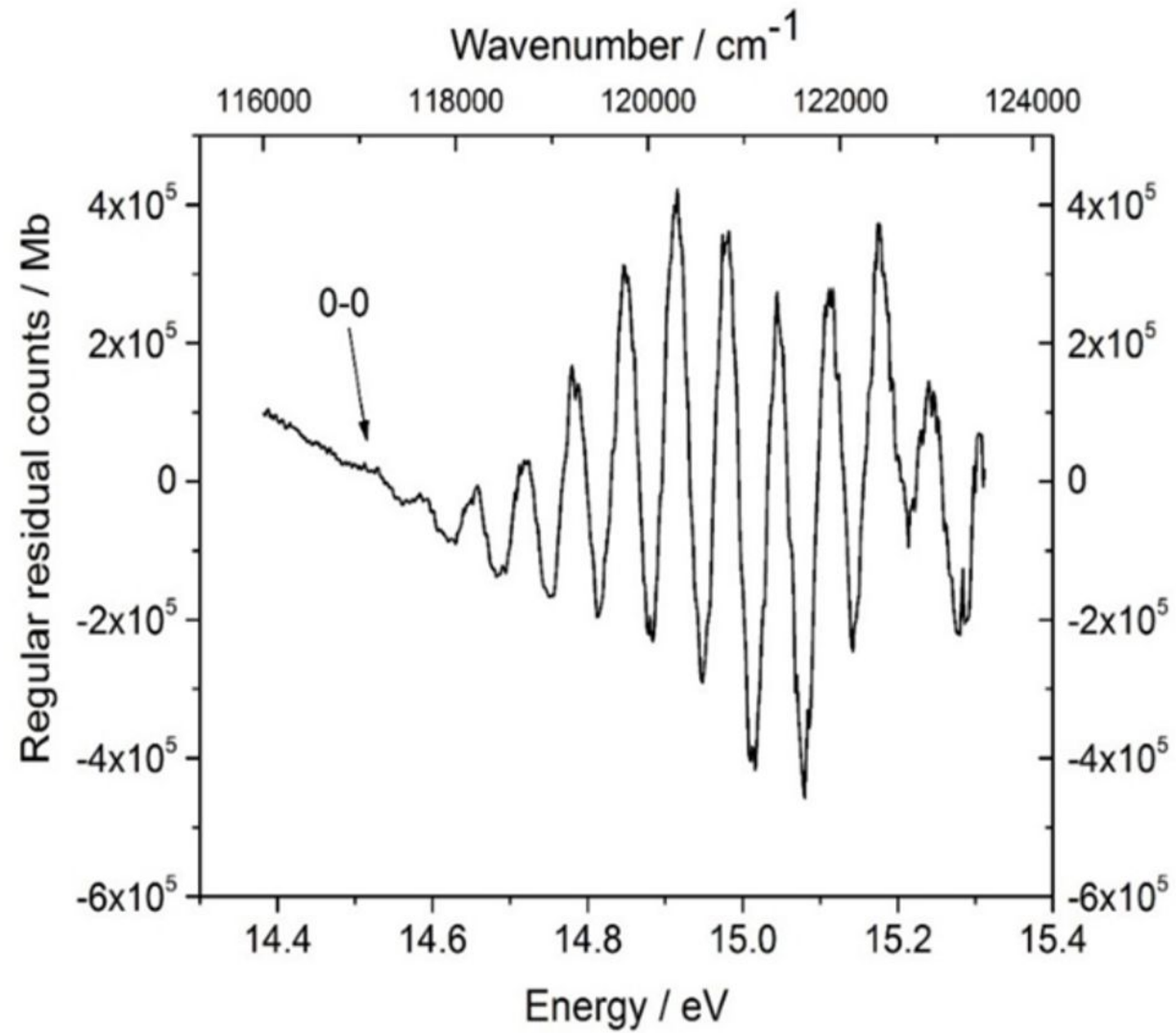

# Palmer et al Fig.12

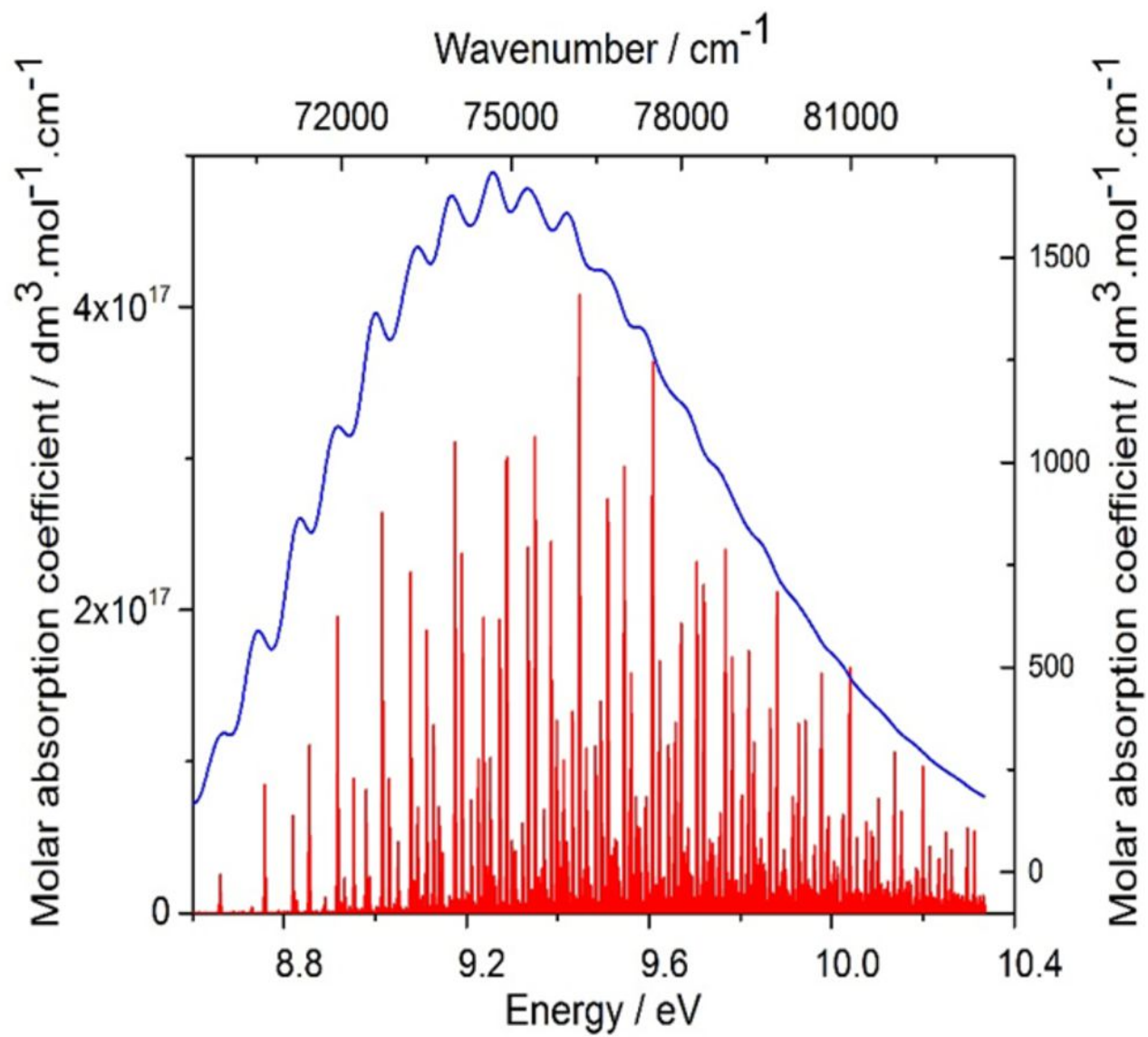

Michael H. Palmer,[1,a] Søren Vrønning Hoffmann,[2,b] Nykola C. Jones,[2,b] Marcello Coreno,[3,b] Monica de Simone,[5,b] Cesare Grazioli,[4,b]

[1] *School of Chemistry, University of Edinburgh, Joseph Black Building, David Brewster Road, Edinburgh EH9 3FJ, Scotland, UK*

[2] *Department of Physics and Astronomy, ISA, Aarhus University, Ny Munkegade 120, DK-8000 Aarhus C, Denmark*

[3] *ISM-CNR, Montelibretti, c/o Laboratorio Elettra, Trieste, Italy*

[4] *CNR-IOM Laboratorio TASC, Trieste, Italy*

[a)] Email: *m.h.palmer@ed.ac.uk*: Telephone: +44 (0) 131 650 4765

[b)] Electronic addresses: vronning@phys.au.dk; nykj@phys.au.dk; marcello.coreno@elettra.eu; desimone@iom.cnr.it; cesare.grazioli@elettra.eu;

**Contents**

SM1. The VUV 'subtraction' process.

SM2. Computational methods expanded. Basis sets.

SM3. Molecular orbital interactions in the excited states of $CH_2F_2$.

SM4. Effect of symmetry and near degeneracy on the excited states.

SM5. Position of Rydberg state functions.

SM6. Conical intersections and avoided crossings.

SM7. Singlet Rydberg states for $CH_2F_2$ using the MRD-CI method

SM8. The harmonic frequencies for the singlet states studied.

SM9. The lowest singlet state of the 3-root CASSCF state-average calculation.

**SM10. Supplementary Material References.**

---------------------------------------------------------------------------------

**SM1. The VUV 'subtraction' process.**

Figure SM1 shows the VUV spectrum used, together with its treatment. A very wide Gaussian function, behaves similar to a ramp, is initially subtracted by use of a set of points touching the VUV curve at appropriate positions. A set of local Gaussian functions is then used in regions of interest to enhance the sharp structure, attributed to Rydberg and possibly valence states. The broad structure is inconsistent with Rydberg states both in style and high intensity, which generally contrasts with the sharp but weak Rydberg state appearance. The latter is expected to be similar to the photoelectron spectrum.

*Figure SM1. The deconvolution process used.*

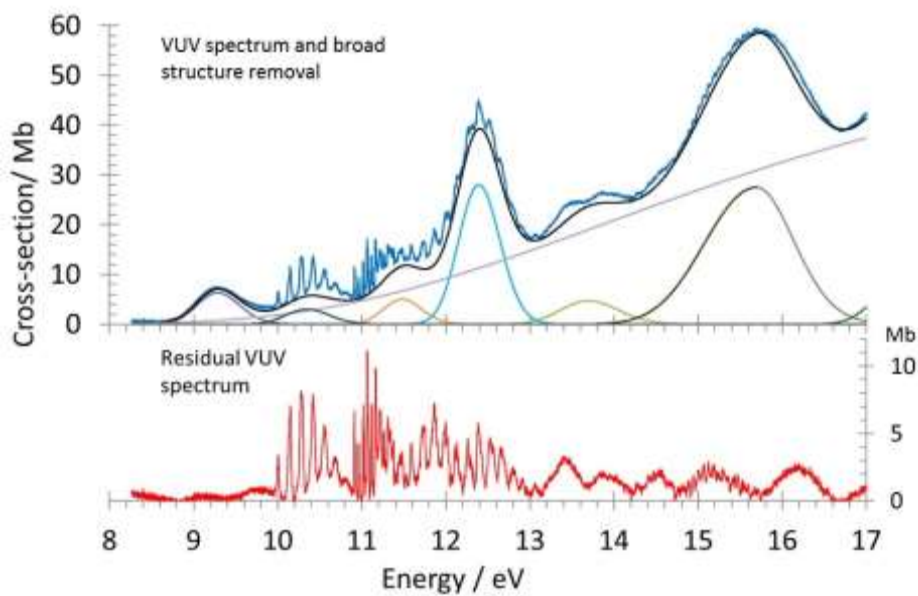

**SM2. Computational methods expanded. Basis sets.** Our results were mainly performed with the optimized Coulomb fitting (second) family of basis sets (DEF2),[27,28] where 'DEF' is an abbreviation for 'default'. Specifically, we used DEF2-QZVP (quadruple zeta valence with polarization) and DEF2-QZVPPD (which contains additional p- and d-diffuse functions). The aug-cc-pVQZ basis set contained [5s4p3d2f] contracted functions for the *H*-atoms, and [6s5p4d3f2g] for the *C* and *F* atoms, including diffuse functions. These will evaluate both the valence states and the lowest Rydberg states correctly; higher Rydberg states require very diffuse exponents, which we have previously used with the triple zeta valence with polarization (TZVP) basis set.[1-4] Here a Rydberg set of functions [4s3p3d3f], was mounted on aug-cc-pVTZ as the background basis. However, this study also required bases showing a clear distinction between valence and Rydberg functions. The balance between basis set quality and bias towards either the ground state or the Rydberg state, has been extensively studied, and the above bases have been recommended for reliable results.[5-7]

**SM3. Molecular orbital interactions in the excited states of *CH$_2$F$_2$*.** The TDDFT method, in common with many related methods, defines the excitation energies (EE) and their associated oscillator strength (f(r)), as the difference in energy between the ground and the excited state, *at the same geometry*. When the equilibrium structures of the X$^1$A$_1$ and (say) X$^1$B$_1$ state are significantly different, these energies require correction to obtain the standard adiabatic value (AEE), in terms of the equilibrium structures of *both* states. The TDDFT procedure generates AEE low by 1 to 2 eV; for these two states, the EE (*G-09*) and AEE are 7.1468 and 8.7111 eV respectively.

**SM4. Effect of symmetry and near degeneracy on the excited states.** The TDDFT and CASSCF methods for calculation of singlet excited states, show mixing of MOs occurs in the leading terms of the wave-functions. These result in two complications; (a) separation into symmetric and antisymmetric combinations of same symmetry; (b) interaction of states of different symmetry in near degenerate situations. Thus, the onset of absorption in the VUV arises from a nearly degenerate pair of states, $1^1$A$_2$ and $1^1$B$_1$, both in the TDDFT and CASSCF methods. Analysis of Band I of the VUV spectrum is made more complex by both these interactions. Important examples, of particular relevance to the VUV Band I, are $1^1$B$_1$ with $2^1$B$_1$ and also $1^1$A$_2$ with $2^1$A$_2$. These states have leading terms:

$1^1$B$_1$:  2b$_1$7a$_1$* - 2b$_1$8a$_1$* - 2b$_1$9a$_1$*
$2^1$B$_1$:  2b$_1$7a$_1$* + 2b$_1$8a$_1$* + 2b$_1$9a$_1$*
$1^1$A$_2$:  2b$_1$5b$_2$* - 2b$_1$6b$_2$* - 2b$_1$7b$_2$*
$2^1$A$_2$:  2b$_1$5b$_2$* + 2b$_1$6b$_2$* + 2b$_1$7b$_2$*

We have studied the potential energy surfaces (PotEnergy) of these same state symmetry interactions by TDDFT methods. It is convenient to use a *CH$_2$F$_2$* structural parameter, angle or bond, to exhibit this. Examples are shown in Figs SM2.1 and SM2.2, where the FCF angle was appropriate. A PotEnergy scan of the lowest state of symmetry ($1^1B_1$ etc) towards the next higher state ($2^1B_1$ etc) of same symmetry, shows the PotEnergy of both states. However, when these two states are similarly scanned in the opposite direction, ie starting with $2^1B_1$ and scanning towards $1^1B_1$, the curves are not identical. The surface exhibits hysteresis, where the region between them is different when the surface is scanned from the higher AEE state to the lower AEE state.

*Figure SM4.1. The potential energy surface for the antisymmetric combination of the $1^1A_2$ and $1^1B_1$ singlet states. This was determined by use of state-averaged CASSCF calculations, using the TZVP basis set with 8 electrons in 8 MOs. The curve gives a close fit to a cubic equation.*

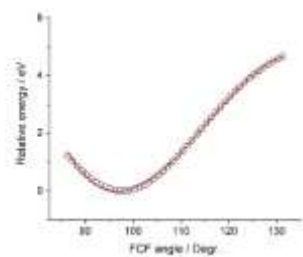

In contrast to the TDDFT program, part of *G-09*, the 'State Average' option in CASSCF calculations, allows mixing of configurations of differing symmetry. This became important, when additional mixing of $1^1B_1$ and $2^1B_1$, with $1^1A_2$ and $2^1A_2$ states became apparent. The surface for the lowest singlet $1^1B_1$ and $1^1A_2$ valence state are close through much of the attractive range, but cross for an FCF angle of 110.5°.

**SM5. Position of Rydberg state functions.** MCSCF calculations were performed where the Rydberg S-type functions were placed on either the ***C***- or both ***H***-, or both ***F***-atoms. Eighteen electrons were processed in 24200 determinants, covering three states of $^1A_1$ symmetry. The equilibrium structures were obtained for the lowest S- and P-Rydberg states, $1^1B_1$ and $1^1A_1$. The choice of placing the Rydberg state functions on the H-atoms gave structures very close to that for the $X^2B_1$ state and close to the $X^1A_1$ state under the same conditions, as shown in Table SM2.3. The alternative choice of the F-atoms gave a much longer C-F bond, and larger HCF angle. Placing the Rydberg functions on the C-atom did not give a realistic structure, since both ***C-H*** and ***C-F*** bonds were lengthened to 1.4 Å and the ***HCF*** angle was reduced to

62°. Overall, the choice of Rydberg functions on the H-atoms seems appropriate. Similar conclusions were found in application to the Rydberg 3S state in *CHF₃*. This supports the views of Edwards and Raymonda that the lowest excited state arises from excitation of the *C-H* bond.[8]

**Table SM5.1. Rydberg state energies and equilibrium structures, using the MCSCF method using the TZVP basis set, augmented by the Rydberg basis set placed on each H-atom.**

| Rydberg state | X$^1$B$_1$ | $^1$B$_1$ | $^1$B$_1$ | $^1$A$_1$ |
|---|---|---|---|---|
| Occupancy | 2b$_1$3S$^b$ | 2b$_1$4s | 2b$_1$3p | 2b$_1$3p |
| Adiabatic IE / eV | 8.6732 | 9.1317 | 7.9265 | 8.9933 |
| C-H bond /Å | 1.1600 | 1.1391 | 1.1712 | 1.1420 |
| C-F bond/ Å | 1.3178 | 1.3291 | 1.3345 | 1.3310 |
| HCH angle/° | 83.70 | 96.2265 | 87.8641 | 99.3940 |
| FCF angle/° | 121.93 | 115.1406 | 113.9785 | 114.9951 |
| HCF angle/° | 111.19 | | | |

**Footnotes to Table SM5.1**

a. MCSCF X$^1$A$_1$ occupancy for a$_1$, b$_1$, b$_2$, a$_2$ orbitals as: [2200, 220, 220, 20] respectively. Number of configuration state functions (CSFs) 42746, generating 157016 determinants.
b. Leading terms [22α0, 2β0, 220, 20]

**SM6. Conical intersections and avoided crossings.** The CASSCF module within the *G-09* suite, determines the position of closest energy approach for two states, while sharing a common structure. An avoided crossing or conical interaction (Conint), is determined by inspection of the wave-function at the common point. We have checked whether an avoided crossing or a conical interaction occurs between the two states 1$^1$B$_1$ and 2$^1$B$_1$, since a Conint would cause major perturbation to the vibrational levels. The *C-H* bond length provides a suitable variable for these potential energy surfaces of *CH₂F₂*. The surface in Fig. SM2.2, shows a crossing of the two nearly touching curves, but only when the molecular parameters have *C-H* 1.204, *C-F* 1.380 (Å) with *HCH* 67.1° and *FCF* 110.5°. These geometric parameters are distant from the minima, and cannot influence the vibrational

*Figure SM6.1. The pair of mixed $1^1A_2$ and $1^1B_1$ singlet states in symmetric and antisymmetric linear combinations, near the curve crossing. The method used is CASSCF state-averaged method for conical interactions. The relative energy scale is arbitrary, but expresses the individual separations of the second and third states from the ground state (root one). The parameters at the crossing are shown, are not greatly different from the equilibrium structures.*

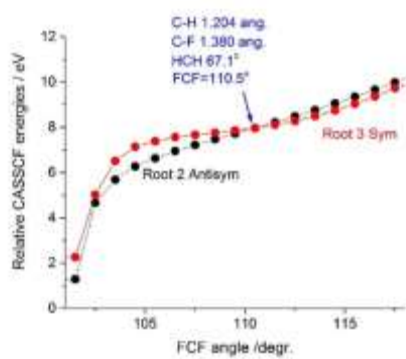

## SM7. Singlet Rydberg states for $CH_2F_2$ using the MRD-CI method

*Figure SM7. The VUV spectrum of $CH_2F_2$, with the MRD-CI vertical excitation energies with their oscillator strengths for the Rydberg states. These states generally show very low f(r), necessitating a separation from the valence state results.*

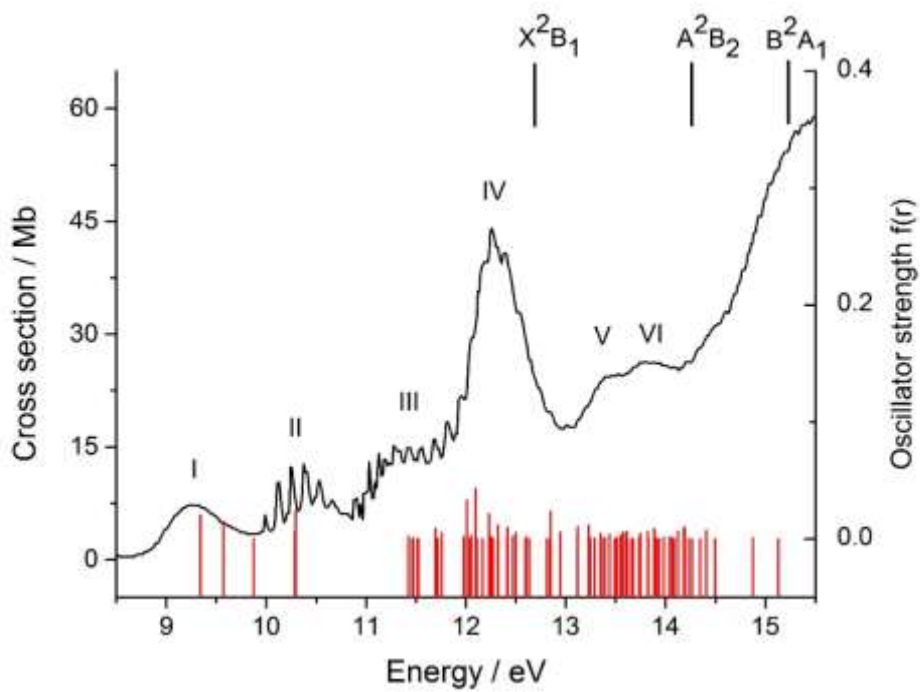

**Table SM7. Selected vertical excitation energies (eV), including the leading configurations and second moments of the charge distribution (a.u.$^2$)**

| Energy / eV | f(r) | Symmetry | Leading configurations | $\langle x^2 \rangle$ | $\langle y^2 \rangle$ | $\langle z^2 \rangle$ |
|---|---|---|---|---|---|---|
| 0.0 | 0.0 | X$^1$A$_1$ | 1-4a$_1^2$;1-2b$_1^2$;1-3b$_2^2$;1a$_2^2$ | -11.6 | -14.2 | -11.7 |
| 9.508523 | 0.024374 | $^1$B$_1$ | 5,7-48 | -27.5 | -23.6 | -29.0 |
| 10.393685 | 0.010825 | $^1$B$_1$ | 6,10-48 | -19.5 | -19.8 | -31.9 |
| 10.597752 | 0.029042 | $^1$A$_1$ | 2b$_1$3b$_1$* | -42.3 | -21.3 | -23.7 |
| 11.773351 | 0.016653 | $^1$B$_2$ | 6,5-82 | -24.7 | -22.1 | -20.9 |
| 11.946601 | 0.001116 | $^1$A$_1$ | 4a$_1$5a$_1$* | -28.5 | -23.6 | -23.8 |
| 12.103246 | 0.084329 | $^1$B$_1$ | 7,8,13-48 | -28.4 | -24.0 | -25.8 |
| 12.635757 | 0.098573 | $^1$A$_1$ | 4a$_1$6a$_1$* | -19.9 | -19.5 | -37.2 |
| 12.695239 | 0.061053 | $^1$B$_2$ | 4-83 | -20.8 | -32.2 | -20.6 |
| 12.814916 | 0.039164 | $^1$B$_1$ | 83-113;8-48 | -22.5 | -33.8 | -22.5 |
| 13.139852 | 0.080144 | $^1$A$_1$ | 2b$_1$5b$_1$* | -34.6 | -19.3 | -23.9 |
| 14.457600 | 0.049249 | $^1$B$_1$ | 4-50,49 | -49.9 | -21.4 | -34.3 |
| 14.800230 | 0.029188 | $^1$B$_1$ | 11-48 | -18.3 | -28.7 | -21.5 |
| 15.140321 | 0.063894 | $^1$B$_2$ | 48-115;51-113 | -27.4 | -25.7 | -24.4 |
| 15.370028 | 0.079545 | $^1$B$_2$ | 9,8-82 | -19.0 | -31.7 | -21.0 |
| 15.676760 | 0.036868 | $^1$B$_2$ | 51-113;48-115 | -33.6 | -19.4 | -27.9 |
| 15.868615 | 0.059742 | $^1$A$_1$ | 3a$_1$5a$_1$* + 3b$_2$6b$_2$* +2b$_1$8b$_1$* | -23.9 | -27.7 | -26.5 |
| 16.001486 | 0.047647 | $^1$B$_1$ | 12-48 | -20.8 | -17.9 | -25.9 |
| 16.067505 | 0.049530 | $^1$B$_1$ | 13-48;84-113 | -29.0 | -27.6 | -28.1 |
| 16.154915 | 0.063668 | $^1$B$_2$ | 6-81;10-82 | -21.5 | -21.0 | -37.6 |
| 16.402301 | 0.050807 | $^1$B$_2$ | 10-82;6-81 | -22.4 | -23.7 | -33.2 |
| 16.469211 | 0.095619 | $^1$A$_1$ | 4a$_1$10a$_1$* - 4a$_1$9a$_1$* | -22.0 | -23.1 | -34.2 |

[a] SCF energy -237.99596 a.u.; orbital occupancy 1a$_1^2$ - 6a$_1^2$, 1b$_1^2$ -2b$_1^2$, 1b$_2^2$ – 4b$_2^2$, 1a$_2^2$

[b] Excitation energies are relative to the $\widetilde{X}$ $^1$A$_1$ ground state CI energy -238.39826 a.u

[c] Singly occupied orbitals except where shown; 97 active orbitals.

**SM8. The harmonic frequencies for the singlet states studied.**

**Table SM8. Comparison of the vibrational modes (1 to 9) and harmonic frequencies (cm$^{-1}$) for the lowest singlet states with the X$^1$A$_1$ state and the Rydberg state ionic core X$^2$B$_1$.**

| State | X$^1$A$_1$ | X$^2$B$_1$ | 1$^1$A$_2$ | 1$^1$B$_1$ | 2$^1$B$_1$ | 1$^1$A$_1$ | 2$^1$A$_1$ | 1$^1$B$_2$ | 3$^1$B$_1$ | 3$^1$A$_1$ | 4$^1$A$_1$ | 4$^1$B$_1$ |
|---|---|---|---|---|---|---|---|---|---|---|---|---|
| 1a$_1$ | 3100 | 2514 | 3158 | 2198 | 2588 | 2402 | 2355 | 2923 | 2450 | 2684 | 2704 | 2589 |
| 2a$_1$ | 1570 | 1292 | 1187 | 1289 | 1131 | 1286 | 1281 | 1180 | 1194 | 1450 | 1391 | 1376 |
| 3a$_1$ | 1150 | 1096 | 660 | 782 | 1021 | 1033 | 1059 | 1090 | 767 | 116 | 1041 | 1149 |
| 4a$_1$ | 540 | 604 | 271 | 562 | 262 | 607 | 603 | 485 | 545 | 468 | 461 | 560 |
| 5a$_2$ | 1297 | 997 | 401 | 1037 | 1063 | 1146 | 1206 | 1062 | 1029 | 1120 | 1177 | 1026 |
| 6b$_1$ | 3172 | 2069 | 3346 | 1892 | 2181 | 2089 | 2052 | 3068 | 2204 | 3085 | 3180 | 2198 |
| 7b$_1$ | 1207 | 618 | 594 | 747 | -622 | 887 | 1119 | 1112 | -840 | 883 | 944 | 351 |
| 8b$_2$ | 1486 | 1512 | 553 | 1486 | 1249 | 1476 | 1471 | 1477 | 1308 | 1167 | 1189 | 1374 |
| 9b$_2$ | 1143 | 1085 | -5935 | 855 | 411 | 1076 | 1064 | 805 | 942 | -1473 | -1446 | 1235 |

**SM9. The lowest singlet state of the 3-root CASSCF state-average calculation.**

The lower singlet state, shown in Fig. SM8, gives a close fit to a cubic surface, and this is marked in red.

*Figure SM9. The potential energy surface for the lowest mixed ($1^1A_2$ - $1^1B_1$) state using state-averaged CASSCF calculations, and the TZVP basis set with 8 electrons in 8 MOs. The curve gives a close fit (shown in red) to a cubic equation. The derived frequencies show more low frequency modes than the $^1B_1$ state in isolation, and this will lead both to a high density of states, and the more closely spaced vibraitonal separations shown in Fig.4 above.*

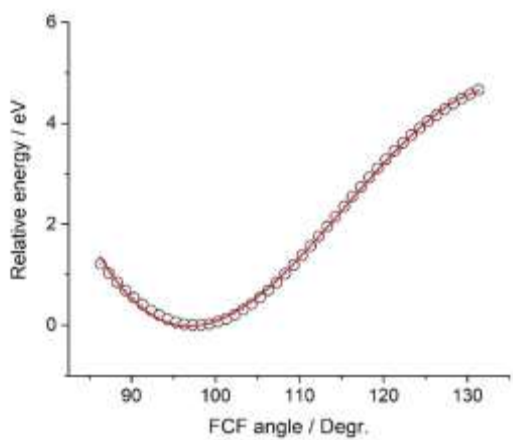

**SM10. Supplementary Material References.**
1. T. H. Dunning, J. Chem. Phys. **55**, 716 (1971)
2. A.D. McLean and G.S. Chandler, J. Chem. Phys. **72**, 5639 (1980).
3. D. E. Woon, and T. H. Dunning, J. Chem. Phys., **100**, 2975 (1994). Gaussian basis sets for use in correlated molecular calculations. IV. Calculation of static electrical response properties